\documentclass{aa}  

\usepackage{graphicx}
\usepackage{txfonts}
\usepackage{hyperref}
\usepackage{orcidlink}
\usepackage{url}
\usepackage[normalem]{ulem}
\usepackage{xcolor}
\hypersetup{
    colorlinks=true,
    linkcolor=blue,
    filecolor=magenta, 
    pdftitle={Overleaf Example}, 
    pdfpagemode=FullScreen,    
    urlcolor=blue,
    citecolor=blue
}
\usepackage[T1]{fontenc}
\defcitealias{farina2022}{F22}
\defcitealias{mazzucchelli2023}{M23}
\defcitealias{Valentini:2021}{V21}
\begin{document}

   %\title{Is GN-z11 powered by a super-Eddington accreting massive black hole?: Extreme value statistics application}
   \title{Is GN-z11 powered by a super-Eddington massive black hole?}

   \subtitle{}
   \authorrunning{Maulik Bhatt et al.}
   \author{Maulik Bhatt  \orcidlink{0000-0002-0287-7485} \fnmsep\thanks{\href{mailto:maulik.bhatt@sns.it}{maulik.bhatt@sns.it}} 
          \inst{1},
          Simona Gallerani\orcidlink{0000-0002-7200-8293}\inst{1},
          Andrea Ferrara \orcidlink{0000-0002-9400-7312}\inst{1},
          Chiara Mazzucchelli \orcidlink{0000-0002-5941-5214}\inst{2},
          Valentina D’Odorico\orcidlink{0000-0003-3693-3091}\inst{1,3,4},
          Milena Valentini\orcidlink{0000-0002-0796-8132}\inst{4,5,6},
          Tommaso Zana\orcidlink{0000-0003-4244-8527}\inst{7},
          Emanuele Paolo Farina\orcidlink{0000-0002-6822-2254}\inst{8},
          \and
          Srija Chakraborty\inst{1}
          %\fnmsep
          %\thanks{Just to show the usage
          %of the elements in the author field}
          }

   \institute{Scuola Normale Superiore, Piazza dei Cavalieri 7, I-56126 Pisa, Italy,
        \and 
        Instituto de Estudios Astrof\'{\i}sicos, Facultad de Ingenier\'{\i}a y Ciencias, Universidad Diego Portales, Avenida Ejercito Libertador 441, Santiago, Chile
        \and 
        INAF – Osservatorio Astronomico di Trieste, via Tiepolo 11, I-34131 Trieste, Italy
        \and
        IFPU - Institute for Fundamental Physics of the Universe, Via Beirut 2, 34014 Trieste, Italy
        \and
        Astronomy Unit, Department of Physics, University of Trieste, via Tiepolo 11, I-34131 Trieste, Italy 
        \and 
        ICSC - Italian Research Center on High Performance Computing, Big Data and Quantum Computing
        \and 
        Dipartimento di Fisica, Sapienza, Università di Roma, Piazzale Aldo Moro 5, 00185 Roma, Italy
        \and
        Gemini Observatory, NSF’s NOIRLab, 670 N A’ohoku Place, Hilo, Hawai'i 96720, USA
             }

   \date{Received January 24, 2024; accepted March 7, 2024}

% \abstract{}{}{}{}{} 
% 5 {} token are mandatory
 
  \abstract
{Observations of $z \sim 6$ quasars powered by supermassive black holes (SMBHs; $M_{\rm BH} \sim 10^{8-10}\, M_\odot$) challenge our current understanding of early black hole (BH) formation and evolution. The advent of the \textit{James Webb Space Telescope} (JWST) has enabled the study of massive BHs (MBHs; $M_{\rm BH}\sim 10^{6-7} \ \mathrm{M}_\odot$) up to $z\sim 11$, thus bridging the properties of $z\sim 6$ quasars to their ancestors.}
{The JWST spectroscopic observations of GN-z11, a well-known $z=10.6$ star-forming galaxy, have been interpreted with the presence of a super-Eddington (Eddington ratio  $\equiv \,\lambda_{\rm Edd}\sim 5.5$) accreting MBH. To test this hypothesis, we used a zoom-in cosmological simulation of galaxy formation and BH co-evolution.} 
{We first tested the simulation results against the observed probability distribution function (PDF) of $\lambda_{\rm Edd}$ found in $z\sim 6$ quasars. Then, in the simulation we selected the BHs that satisfy the following criteria:  (a) $10 < z < 11 $, (b)  $M_{\rm BH} > 10^6 \ \mathrm{M}_\odot$. Next, we applied the extreme value statistics to the PDF of $\lambda_{\rm Edd}$ resulting from the simulation.}
{We find that the probability of observing a $z\sim 10-11$ MBH accreting with $\lambda_{\rm Edd} \sim 5.5$ in the volume surveyed by JWST is very low ($<0.2\%$). We compared our predictions with those in the literature, and discussed the main limitations of our work.}
{Our simulation cannot explain the JWST observations of GN-z11. This might be due to: (i) poor resolution and statistics in simulations, (ii) simplistic sub-grid models (e.g. BH accretion and seeding), (iii) uncertainties in the data analysis and interpretation.}
%, {\bf associated e.g. to the identification of high-z AGN via the presence of broad and/or highly ionized emission lines.}}


   \maketitle
%
%-------------------------------------------------------------------

\section{Introduction}
Observations have shown that the most distant quasars known so far \citep[redshift $z = 6-7.5$; see a recent review by][]{fan2023} are powered by supermassive black holes (SMBHs) with masses $\sim 10^{8-10}~\rm M_{\odot}$ \citep[e.g.][]{banados2018,yang2020,wang2021,farina2022, mazzucchelli2023,yang2023}. The existence of these gigantic black holes (BHs) presents a puzzle for current theoretical models of BH formation and evolution that is still lacking several pieces, for example, knowledge about the SMBH seed nature and mass and their ability to  grow fast enough to assemble an SMBH in less than 1 Gyr, the age of the Universe at $z\sim 6$. Studying massive black holes (MBHs) of $\sim 10^{6-7} \ \mathrm{M}_\odot $ at redshift  $z>9$ is essential in order to take a step forward in this field \citep{inayoshi2020,volonteri2012}. 

The most recent \textit{James Webb Space Telescope} (JWST) observations have revealed the presence of accreting ($L_{\rm bol}\sim 10^{44-47}~\rm L_\odot$) MBHs in galaxies up to $z\sim 10-11$ \citep[e.g.][]{bosman2023,greene2023,goulding2023,furtak2023,kokorev2023,larson2023,maiolino2023a}. These data are powerful tools for constraining different seeding scenarios \citep[e.g.][]{jeon2024,trinca2022,trinca2023,pacucci2023, schneider2023}. A lot of interest in particular has been aroused by the detection of a vigorously accreting MBH in GN-z11, a well-known star-forming galaxy at $z=10.6$ \citep{maiolino2023a}.
%Super massive black holes (SMBH) of $ \sim 10^8-10^9 \ \mathrm{M}_\odot $  are believed to power the most luminous quasars at $ z > 6$. However, understanding the astrophysical origin of these SMBHs still is a theoretical challenge. Exploring intermediate mass black holes of $  \sim 10^6 - 10^7 \mathrm{M}_\odot $ at redshift  $z>9$ can be essential to understand the growth of the black holes in early universe \citep{inayoshi2020,volonteri2012}, and to explain the possible origin of the super massive black holes powering $z\sim 6-7.5$ quasars \citep[see the review by][]{fan2023}. Recent claim of a small and vigorous black hole in GN-z11 can possibly be the seed of SMBH at z $\sim6-7$ \citep{maiolino2023a}.
 %in red shift range $6 < z < 7.5$ (\cite{2021NatRP...3..732V}) detected in various surveys such as SDSS (\cite{2001AJ....122.2833F}, \cite{2003AJ....125.1649F}, \cite{2006AJ....131.1203F}), Canada-France High-z Quasar Survey (\cite{2005ApJ...633..630W}, \cite{2009AJ....137.3541W}, \cite{2010AJ....139..906W}), Visible and Infrared Survey Telescope for Astronomy Kilo-degree INfrared Galaxy (\cite{2013ApJ...779...24V}, \cite{2015MNRAS.453.2259V})

The galaxy GN-z11 was first identified by \citet{bouwens2010} and \citet{oesch2016}. Recently, the JWST Advanced Deep Extragalactic Survey \citep[JADES;][]{eisenstein2023} provided follow-up observations by means of NIRSpec spectroscopy \citep{bunker2023} and NIRCam imaging \citep{tacchella2023}. These observations suggest that GN-z11 is an active galactic nucleus (AGN). The evidences in favour of this interpretation include the followings \citep{maiolino2023a}: (i) high ionisation transitions (e.g. [NeIV]$\lambda2422,2244$), which are commonly observed in AGN \citep{terao2022,lefevre2019} and are considered as tracers of AGN activity since they require high photon energies ($E_{\rm \nu} > 63.5 \ \mathrm{eV})$ that are not easily produced by stars; (ii) {\bf $\rm NIII]$} multiplet, indicating  gas densities $( > 10^{10} \ \rm cm^{-3})$ typically found in the broad-line regions (BLRs) of AGN; (iii) a very deep equivalent width $(\mathrm{EW_{rest} \sim 5 \AA})$ and a blueshifted absorption trough of a CIV doublet, suggesting the presence of outflows with velocities ($\sim 800-1000 \ \mathrm{km \ s^{-1}}$) commonly observed in mini-broad absorption line (BAL) quasars \citep{rodriguez2009phd}; 
(iv) a potentially broad profile of the $
\rm HeII \lambda
1640$ line that could come from the inner region of the BLR but could also trace the presence of a Wolf–Rayet (WR) stellar population; (v) clear detection of an MgII doublet hinting at the presence of an accreting MBH with mass $\sim 1.2  \times 10^6 \ \mathrm{M}_\odot$ and a surprisingly large Eddington ratio ($\lambda_{\rm Edd}=L_{\rm bol}/L_{\rm Edd}\sim 5.5 \pm 2$).
%estimated assuming the local virial relations \citep{vestergaard2009,buendia-rios2023}. The inferred 
%Eddington ratio is 5.5 with uncertainty of a factor of at least 2 \citep{maiolino2023a}.
%that super-Eddington accretion in well-formed, evolved galaxies is an attractive route to the formation of MBHs. Lupi 2016

%Lupi 2016, starting from light seeds (10-100 Msun) you can form a MBH (1.e3-1.e5) at z~10-15. Zhu, starting from light seeds (10-100 Msun) at z~20 you cannot form a SMBH (1.e8-1.e10) at z~6. 

%Smith PopIII cannot make SMBHs at z~6 but they do not have slim disk

%Pezzulli (feedback+slim disk) PopIII can make

% Madau PopIII can make but they do not have feedback

%Pacucci 2017 (Natarajan, Volonteri) 

%The high accretion rate of GN-z11 has important implications on the nature of the seeds that give birth to massive black holes in the early Universe. In fact, BHs originated from PopIII stars can explain the existence of $z\sim 6$ SMBHs ($z\sim 10$ MBHs), only if they accrete super-Eddington ($f_{\rm Edd}=\dot{M}/\dot{M}_{Edd}>1$) for a sufficiently long interval of time  

The high-accretion rate of GN-z11 provides important constraints for theoretical models predicting the mass and the formation epoch of early SMBH seeds \citep{inayoshi2020,volonteri2021,latif2016}. The various possibilities include: (1) light seeds ($\sim$10$^{1-2}$~$\rm M_{\odot}$) from Pop III stars \citep[z$\sim$20-30,][]{madau2001}; (2) intermediate seeds ($\sim$10$^{3-4}$~$\rm M_{\odot}$) from the collapse of supermassive stars or runaway collisions in dense nuclear star clusters; \citep[z$\sim$10-20,][]{devecchi2012,greene2020,kroupa2020} (3) heavy seeds ($\sim$10$^{5-6}$~$\rm M_{\odot}$) produced by  the collapse of metal-free gas in atomic-cooling (virial temperature $T_{vir}>10^4$~K) halos exposed to a strong Lyman-Werner radiation (z$\sim$10-20), the so-called direct collapse BHs \citep[DCBHs;][]{ferrara2014,haehnelt1993,mayer2019}. 

Of the above scenarios, the most favourable to explain the existence of SMBHs at high-$z$ is still unclear. On the one hand, the DCBH scenario has the advantage that heavy seeds can grow at a mild pace, namely at sub-Eddington rates ($f_{\rm Edd}=\dot{M}/\dot{M}_{\rm Edd}<1$). However, the conditions for the formation of DCBH are not easily satisfied.  
On the other hand, light/intermediate seeds can form in less extreme circumstances, but they require  sustained  ($f_{\rm Edd}>1$) accretion rates for prolonged ($\sim \rm Gyr$) intervals of time. Recent observations of $10^{7-8} \ \rm M_{\odot}$ BHs at $z\sim 7-8$ \citep[e.g.][]{larson2023,kokorev2023} can be explained with both scenarios. The observed properties of GN-z11 can be reproduced by super-Eddington slim-disc accreting BHs descending from light and heavy seeds \citep{schneider2023}; the more extreme case of UHZ-1 at $z\sim 10$ \citep{goulding2023} lends support to the heavy seeding channel. 

Semi-analytical works that support light seed scenarios are based on radiatively inefficient `slim-disc' models in which super- or hyper-Eddington accretion \citep{sadowski2009} can occur \citep[e.g.][]{madau2014, pezzulli2016, pezzulli2017,trinca2022}. However, it is still not completely understood how much these results depend on the assumptions adopted. To start with, \citet{madau2014} assumed that, in super- or hyper-Eddington accretion regimes, gas can flow towards the centre BH almost non-affected by feedback processes. However, radiative feedback in radiatively efficient `thin disk' models \citep[e.g.][]{pacucci2015, orofino2018} has been shown to modify the accretion flow onto the BH, either decreasing the accretion to sub-Eddington rates or making $f_{\rm Edd}>1$ episodes intermittent. In particular, \cite{pacucci2017} have shown that very efficient and prolonged large accretion rates only occur in $M_{\rm BH}>10^4~M_{\odot}$. For what concerns the \citet{pezzulli2017} results, the adoption of $f_{\rm Edd}>500$ values, which goes beyond the applicability of the adopted `slim-disc' recipe, may overestimate the accretion rate of light seeds and therefore the final mass of the formed SMBH. 

%However, these models assume that, in these regimes, gas can flow toward the center BH almost non-affected by feedback processes \citep{madau2014} or allow for $f_{\rm Edd}>500$, values that go beyond the applicability of the adopted "slim disk" recipe \citep{pezzulli2017}. Radiative feedback in radiatively-efficient "thin disk" models \citep[e.g.][]{pacucci2015, orofino2018} has been shown to modify the accretion flow onto the BH both decreasing the accretion to sub-Eddington rates or making $f_{\rm Edd}>1$ episodes intermittent. In particular, \cite{pacucci2017} have shown that very efficient and prolonged large accretion rates only occur in $M_{\rm BH}>10^4~M_{\odot}$. 

The hydrodynamical simulations developed so far have provided further important information about this issue. \cite{lupi2016} produced high-resolution ($< 1$pc) simulations of an isolated disc in which the radiatively inefficient supercritical accretion of light seeds occurs in the high density environment of gaseous circumnuclear discs. These authors find that 10-100 M$_{\odot}$ BHs can increase their mass within a few million years by between one and up to three orders of magnitude, depending on the resolution adopted (the higher the resolution, the lower the final mass of the grown seed). Less efficient growth of light seeds has been found by \cite{smith2018}, who followed the evolution of Bondi accreting BHs from Pop III stars in high-resolution cosmological simulations, finding a much smaller ($<$10\% of their initial mass) average mass increase. Similar conclusions have been drawn by a comprehensive study performed by \cite{zhu2022}, who considered various recipes for BH seeding, accretion models, and feedback processes. Even in the most optimistic conditions (radiatively inefficient, super Eddington accretion) and despite spending a substantial fraction of time ($\sim 100$~Myr) in super-critical accretion, light seeds ($M< 10^3 \ \mathrm{M}_\odot$) cannot reproduce $>10^7 \ \mathrm{M}_\odot $ ($>10^5 \ \mathrm{M}_\odot $) BHs at $z\sim 6$ ($z\sim 11$).   

%In particular, Smith et al. (2018) followed the growth of over 15000 BHs from Pop III stars in the Renaissance simulations with ultra-high resolutions down to mDM = 127M /ℎ, and they found inefficient growth for most of the BHs, with an average mass increase less than 10%.

%When accounting for , the accretion is very efficient and proceeds continuously with large accretion rates (fEdd >>1    
%accretion is probably episodic, if it has been occurring for the previ-
%ous   100 Myr, then the black hole could have potentially originated
%even from a stellar mass seed at z 12-15.

Current investigations have not definitively clarified whether light seeds can provide a valuable route for the formation of high-$z$ MBHs. Furthermore, high-resolution (computationally expensive) hydrodynamical simulations are required to properly follow the growth of light seeds. Several works \citep[see Table 1 in][]{habouzit2021} have thus opted to assume the DCBH scenario as a seeding prescription. In this work, we adopted the cosmological zoom-in hydrodynamical simulations developed by \citet[][hereafter \citetalias{Valentini:2021}]{Valentini:2021}. These simulations can reproduce the main properties of the most luminous $z\sim 6$ quasars (see Tables 2 and 3 in \citetalias{Valentini:2021}), namely the mass of their central SMBHs ($\sim 10^9~\rm M_{\odot}$) suggested by observations of the CIV and MgII lines \citep[e.g.][]{farina2022,mazzucchelli2023}, their high star formation rates (SFRs; i.e. $\sim 100\rm ~M_{\odot}~yr^{-1}$), and their large masses ($\sim 10^9~\rm M_{\odot}$) of molecular gas suggested by ALMA observations \citep[e.g.][]{wang2016,venemans2017,venemans2020,gallerani2017,decarli2018,decarli2022}. In this work, we investigate the predictions of \citetalias{Valentini:2021} simulations in terms of the accretion properties that characterise $z\sim 10-11$ BHs as massive as the one hosted in GN-z11 ($M_{\rm BH}\gtrsim 10^{6}~\rm M_{\odot}$). 

%These unprecedented observations raise question about the possible extreme values of super-Eddington accretion rate of massive black holes in the early universe. 

%If the GN-z11 black hole has been accreting at the super-Eddington rate also at previous times then a stellar black hole seed formed by pop-III stars at z$\sim 12-15$ can possibly explain the estimated black hole mass \citep{lupi2016} However, zoom-in hydrodynamical simulation by \cite{zhu2022} reported that low mass BH seed $< 10^3 \mathrm{M}_\odot$ did not able to reproduce $10^8 \mathrm{M}_\odot $ at $z\sim 6$ or $10^6 \mathrm{M}_\odot $ at $z\sim 11$ even with super-Eddington accretion and the final BH mass in the case with super/hyper Eddington accretion is lower than the Eddington limited case. Additionally, some models have found it unlikely to sustain super-Eddington accretion for a long period \citep{smith2018,pacucci2017} indicating the observed accretion rate of GN-z11 can be episodic. This episodic scenario requires massive seed formed from Direct Collapse Black Hole \citep[DCBH, see][]{latif2013} or nuclear clusters \citep{devecchi2009} accreting with the Eddington or sub-Eddington rate to produce BH of mass $\sim10^6 \mathrm{M}_\odot$ at z=10.6. 

%Therefore, understanding the possible extreme value predictions of the episodic super-Eddington accretion for the early massive black holes from the numerical simulations can provide a strong link to the black hole seed formation \citep{bellovary2011,latif2016,dijkstra2014} and earlier phases of Black Hole (BH) growth and evolution \citep{volonteri2008}. 

In particular, to test whether the \citetalias{Valentini:2021} simulations can reproduce the $\lambda_{\rm Edd}\sim 5.5$ found in GN-z11, we adopted the extreme value statistics \citep[EVS;][]{gumbel1958statistics,kotz2000extreme}. The EVS have been applied to a wide range of topics in cosmology \citep{lovell2023,harrison2011,colombi2011,chongchitnan2012,davis2011,waizmann2011,mikelsons2009,chongchitnan2021} and can be used to calculate the probability of randomly extracting the largest (or smallest) value from an underlying distribution.

This paper is organised as follows: In Sec. ~\ref{sec:simulations}, we describe the adopted numerical simulations. We then test the accretion model adopted in our simulations in Sec. ~\ref{sec:comparison} by comparing the simulated Eddington ratio probability distribution function (PDF) with the largest sample of sources for which this quantity has been measured so far. In Sec. ~\ref{sec:EVS_theory}, we compute the exact EVS PDF of the Eddington ratio for GN-z11-like MBH from the simulations. Finally, we discuss our results and draw our conclusions in Sec. ~\ref{sec:conclusion}.
Throughout the paper we have considered a flat $\Lambda $CDM cosmology with the following cosmological parameter values: baryon density $\Omega_{b} h^{2} = 0.0224 $, DM density $\Omega_{\rm dm} h^2 = 0.12$, Hubble constant $ H_{0} = 67.74 \ \mathrm{km \ s^{-1} \ Mpc^{-1}}= 100 h$, and the late-time fluctuation amplitude parameter $\sigma_{8} = 0.826$ \citep{planck2020}.
\section{Simulations}
\label{sec:simulations}
We used the hydrodynamical cosmological zoom-in simulations developed by \citetalias{Valentini:2021}. We summarise the main features of this model below. In particular, we considered the {\it AGN fiducial} run featuring thermal AGN feedback as our fiducial model, and we refer the reader to \citetalias{Valentini:2021} for more details. 
%In this section, we describe briefly the cosmological hydrodynamical simulations used in this work and physical processes important for this work. We chose a simulation run $AGN\_fid$ from the suites introduced by \citet{Valentini:2021} -- hereafter \citetalias{Valentini:2021}.
The \citetalias{Valentini:2021} simulation adopted in this work was performed with a non-public version of the TreePM (particle mesh) and SPH (smoothed particle hydrodynamics) code GADGET-3 \citep{valentini2017,valentini2019,valentini2020}, an upgraded version of the public GADGET-2 code \citep{springel2005}.
%The \citetalias{Valentini:2021} simulations are performed with the TreePM (particle mesh) and SPH (Smoothed Particles Hydrodynamics) code GADGET-3
 %The simulation follows the evolution of a $\sim 10^{12}$~M$_{\odot}$ halo at $z= 6$. 

%As AGN feedback is very important for accretion we consider AGN fiducial run from \citetalias{Valentini:2021} which includes the thermal feedback as our fiducial model for this work.

%\begin{enumerate}
 %   \item[$\bullet$] AGN fiducial: our fiducial model, featuring thermal AGN feedback;
  %  \item[$\bullet$] \BHsnoFB{}: a control run, analogous to AGN\_fid,  in which BHs and their accretion are included, but AGN feedback is turned off;
%\end{enumerate}
%and the following runs from %\citetalias{Barai2018}:
%\begin{enumerate}
 %    \item[$\bullet$] \AGNcone{}: in which the kinetic feedback is distributed in a bi-cone with and half-opening angle of ${45}$~\textdegree.
  %  \item[$\bullet$] \AGNsphere{}: featuring isotropic, kinetic AGN feedback;
%\end{enumerate}

%In the following sub-sections, we briefly summarize some important features of the simulation relevant for our study. For more details about the simulation refer to the \cite{Valentini:2021}.

%\subsection{Valentini et al. 2021 models: AGN fiducial}
%\label{sec:Valentini_sims}
\subsection{Initial conditions and resolution}
\label{subsec:Valentini_ICs}

The software MUSIC\footnote{MUSIC stands for Multiscale Initial Conditions for Cosmological Simulations: \url{https://bitbucket.org/ohahn/music}.} \cite{hahn2011} was used to generate the initial conditions, assuming a $\Lambda$CDM cosmology. First, a dark matter (DM)-only simulation with a mass resolution of DM particles of $9.4\times 10^8~ \mathrm{M}_\odot$ in a comoving volume of $(148~{\rm Mpc})^3$ was run starting from $z= 100$ down to $z= 6$. Then, a halo as massive as $M_{\rm halo}=1.12 \times 10^{12}~ \mathrm{M}_\odot$ at $z= 6$ was selected for a zoom-in procedure in order to run the full hydrodynamical simulation. In the zoom-in region, the highest resolution particles have a mass of $m_{\rm DM}=1.55\times 10^6 \ \mathrm{M}_\odot$ and $m_{\rm gas}=2.89 \times 10^5 \ \mathrm{M}_\odot$. The gravitational softening lengths are\footnote{A letter \emph{c} before the corresponding unit refers to comoving distances (e.g. ckpc)}%\footnote{We use the following convention when indicating distances: a letter \emph{c} before the corresponding unit refers to \emph{comoving} distances (e.g. ckpc), while the letter \emph{p} refers to \emph{physical} units (e.g. pkpc). When not explicitly stated, we refer to physical distances.} 
$\epsilon_{\rm DM}=0.72$~ckpc and $\epsilon_{\rm bar}=0.41$~ckpc for the DM and the baryon particles, respectively.

\subsection{Sub-resolution physics}
%\begin{enumerate}
%\item[$\bullet$]\ \textbf{Cooling, star formation and stellar feedback{}}: the multiphase interstellar medium (ISM) is described by means of the MUlti Phase Particle Integrator (MUPPI) sub-resolution model \citep{Murante2010, Murante2015, Valentini2017, Valentini2019}. It features metal lines cooling, an \HH-based star formation, thermal and kinetic stellar feedback, the presence of an UV background, and the \citet{Tornatore_2007} model for chemical evolution. 
\subsubsection{Black hole seeding} 
The DM halos exceeding the threshold mass $M_{\mathrm {DM}}=1.48 \times 10^9~\mathrm{M}_\odot$ were seeded with a BH of mass $M_{\mathrm{BH, seed}} = 1.48 \times 10^5~\mathrm{M}_\odot$ if no BH was already seeded in previous time steps. This value of seed mass mimics the DCBH scenario.
%BHs are treated as collisionless sink particles. Seeds of mass $M_{\mathrm{BH, seed}} = 1.48 \times 10^5~\mathrm{M}_\odot$ are implanted in DM halos with mass exceeding $M_{\mathrm {DM, seed}}=1.48 \times 10^9~\mathrm{M}_\odot$. 
%BH repositioning (or \emph{pinning}) is implemented, in order to prevent BHs from wandering from the centre of the halo in which they reside: at each time-step BHs are shifted towards the position of minimum gravitational potential within their softening length \citep[as also done in e.g.][]{2009MNRAS.398...53B, 2015MNRAS.446..521S, 2017MNRAS.465.3291W, 2018MNRAS.473.4077P}.
%\subsubsection{BH growth : Gas accretion on BHs and Merger}
\subsubsection{Black hole growth} 
Black holes grow due to gas accretion and merger with other BHs. The gas accretion was modelled assuming a Bondi–Hoyle–Lyttleton accretion solution \citep{bondi1952,hoyle1939,bondi1944}:
\begin{equation}\label{eq-Mdot-Bondi}
\dot{M}_{\rm bondi}= \frac{4\pi G^2M_{\rm BH}^2 \langle\rho_{\rm gas}\rangle}{(\langle c_{\rm s} \rangle^2+\langle v_{\rm BH}\rangle^2)^{3/2}},
\end{equation}
where $M_{\rm BH}$ is the mass of the BH, $\rho_{gas}$ is the gas density, $G$ is the gravitational constant, $c_s$ is the sound speed, and $v_{\rm BH}$ is the velocity of the BH relative to the gas. All the quantities of the gas particles are calculated within the BH smoothing length using kernel-weighted contributions.
The BH accretion rate was capped to the Eddington value. A small fraction, $\epsilon_{r} \ \dot{M}_{\rm accr}$, of the accreted mass is converted into radiation; thus, the actual growth rate of the BH mass can be written as
\begin{equation}\label{eq:BH_mass_growth}
    \dot{M}_{\rm BH} = ( 1 - \epsilon_{\rm r}) \ \dot{M}_{\rm accr},
\end{equation}
where $\epsilon_{r}$ =0.03 is the radiation efficiency \citep{sadowski2017}. The BHs instantaneously merged if their distance became smaller than twice their gravitational smoothing length and if their relative velocity was smaller than 0.5 $c_{\rm s}$ ($v_{\rm BH-BH} < 0.5 \ \langle c_{\rm s}\rangle$). The position of the resulting BH is the position of the most massive one between the two progenitor BHs.
%besides BH-BH mergers, black holes are also allowed to grow via gas accretion, as described by the classical Bondi-Hoyle-Lyttleton (BHL) model \citep{Hoyle1939, Bondi_Hoyle, Bondi52, Edgar_2004}:
%\begin{equation}  
%\label{eq-Mdot-Bondi} 
%\dot{M}_{\rm Bondi} = \frac{4 \pi G^2 M_{\rm BH}^2 \rho}{ \left(c_{\rm s}^2 + v^2\right) ^ {3/2}} , 
%\end{equation}
%where $G$ is the gravitational constant, $M_{\rm BH}$ is the BH mass, $\rho$ is the gas density, $c_{\rm s}$ is the sound speed, and $v$ is the velocity of the BH relative to the gas. These quantities are evaluated by averaging over the SPH gas particles within the BH smoothing length, with kernel-weighted contributions. 
%Eq. \ref{eq-Mdot-Bondi} is used to estimate the contribution to the accretion rate from the cold and hot phase of the ISM, separately \citep{Steinborn2015, Valentini2020}. Accretion from the cold gas is reduced by taking into account its angular momentum \citep[see][for details]{Valentini2020}. 

\subsubsection{Active galactic nucleus feedback} 
For what concerns AGN feedback, a fraction $\epsilon_{\rm f}$   of the bolometric luminosity $L_{\rm bol} = \epsilon_{\rm r} \dot{M}_{\rm BH}c^2$ is distributed to the gas particles thermally and isotropically within the BH smoothing volume \citep{valentini2020}. \citetalias{Valentini:2021} adopt $\epsilon_{\rm f} = 10^{-4}$, tuned to match the normalisation of the black hole to stellar mass relation at $z=6$.
\section{Eddington ratio predictions against $z\sim 6$ data}
\label{sec:comparison}
\begin{figure*}
\begin{center}
\includegraphics[width=0.98\textwidth]{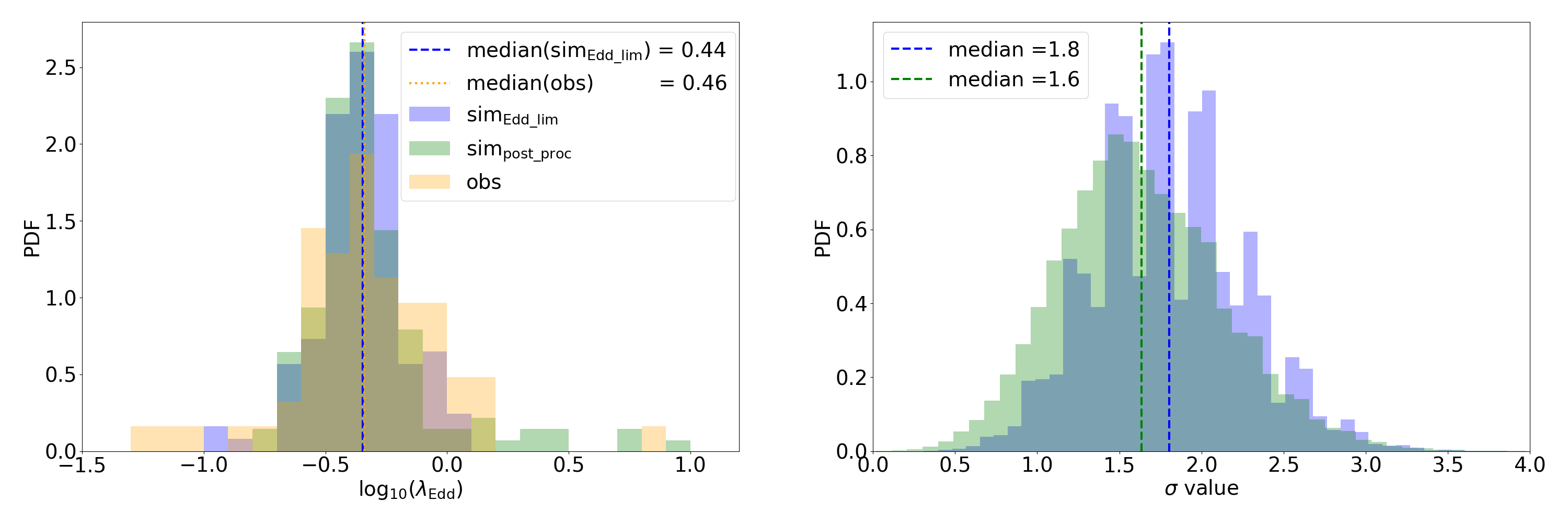}
\caption{Comparison between observations of $z\sim 6$ quasars and simulations. \textsl{Left panel}: 
%Comparison between one realisation of re-sampled (see Sec. \ref{sec:comparison}) Eddington ratio values obtained from the Eddington-limited accretion model (\citetalias{Valentini:2021}; blue histogram), Eddington ratio values obtained in post processing from \citetalias{Valentini:2021} using Eq. \ref{eq-Mdot-Bondi} (green histogram) and 
The orange histogram shows the probability distribution function of Eddington ratios measured from the literature sample of quasars at 6<z<7.5. This is compared with Eddington ratio values obtained from the Eddington-limited accretion model (\citetalias{Valentini:2021}; blue histogram), and with Eddington ratio values obtained in post processing from \citetalias{Valentini:2021} using Eq. \ref{eq-Mdot-Bondi} (green histogram).
%Comparison between one realisation of re-sampled (see Sec. \ref{sec:comparison}) Eddington ratio values obtained from the Eddington-limited accretion model (\citetalias{Valentini:2021}; blue histogram), Eddington ratio values obtained in post processing from \citetalias{Valentini:2021} using Eq. \ref{eq-Mdot-Bondi} (green histogram) and Eddington ratio values measured from a literature sample of quasars at 6<z<7.5 (orange histogram). The  orange (blue) vertical dotted (dashed) line shows the median of the observations (Eddington ratio values from the Eddington-limited accretion model by \citetalias{Valentini:2021}). 
\textsl{Right panel}: Distribution of the $\sigma$ values obtained by applying the two-sample K-S test on $N_{\rm re-samp} = 10^6$ number of resamples. The blue (green) PDF refers to the $\sigma$ distribution corresponding to the original (post-processed) values of $\lambda_{\rm Edd}$ from the \citetalias{Valentini:2021} simulation. The vertical blue and green dashed lines show the median of the blue and green distributions, respectively.}
%\caption{\textsl{Left panel shows comparison between one realisation of the Eddington ratio with Eddington limited accretion from \citetalias{Valentini:2021} in blue shaded region, one realisation of the Eddington ratio calculated in the post processing from \citetalias{Valentini:2021} using Equation \ref{eq-Mdot-Bondi} in orange shaded region and literature sample for the quasars within $6<z<7.5$ in green shaded region. Right panel shows the $\sigma$ distribution obtained from two sample K-S test with $10^6$ number of resamples.  }}
\label{lamda_sub-edd_comp_MV_Farina_S11}
\end{center}
\end{figure*}

In this section, we compare the Eddington ratios $\lambda_{\rm Edd}$ measured for $z\sim 6-7.5$ quasars with the 
%$\left( \lambda_{\rm Edd} = \frac{L_{\rm bol}}{L_{\rm Edd}}\right)$ 
results from the %mass and luminosity controlled Eddington ratio PDF from 
\citetalias{Valentini:2021} simulation. 
%samples to verify the robustness of  considered simulation. 
The Eddington ratio is defined as $\lambda_{\rm Edd} = {L_{\rm bol}}/{L_{\rm Edd}}$, namely the ratio between the bolometric luminosity of a quasar and its Eddington luminosity that, under the assumption of hydrostatic equilibrium and pure ionised hydrogen, can be written as \citep{eddington1926book} %where Eddington luminosity is the theoretical maximum luminosity emitted by quasar when gravity and radiation pressure are in equilibrium in a spherical geometry and can typically be written assuming hydro-static equilibrium of pure ionized hydrogen as follows :
\begin{equation} 
\label{eq-L_Edd} 
L_{\rm Edd} = \frac{4\pi G \ M_{\rm BH} \ m_{\rm p} \ c }{\sigma_{\rm T}},
\end{equation}
where $m_{\rm p}$ is the mass of a proton and $\sigma_{\rm T}$ is the Thomson scattering cross-section.

For what concerns the observed $\lambda_{\rm Edd}$ values, we considered both the sample of 38 quasars by \citet[][hereafter \citetalias{farina2022}]{farina2022} and the sample of 42 quasars by \citet[][hereafter \citetalias{mazzucchelli2023}]{mazzucchelli2023}, the latter includes the XQR-30 sample of VLT-XSHOOTER observations \citep{dodorico2023}. Of the 80 quasars in the combined sample, 18 of them are present in both samples. In these cases, we considered the data characterised by the smallest error. After removing duplicates, we had 62 bright quasars at 5.8 $<$ z $<$ 7.5. Hereafter,  we refer to this combined sample as the `literature sample'. %This contains 68 bright quasars at 5.8 $<$ z $<$ 7.5, from the XQR-30 Large Program \cite{dodorico2023} of VLT/X–shooter observations. 

\citetalias{farina2022} and \citetalias{mazzucchelli2023} provide the $\lambda_{\rm Edd}$ values resulting from the analysis of the MgII and CIV emission lines using different methods (MgII line:  \citealt{vestergaard2009},  \citealt{shen2011};  CIV line: \citealt{vestergaard2006},  \citealt{coatman2017}). When the MgII line was available, we used the \cite{shen2011} method since it provides the smallest errors \citep[see also][for a comparison among different virial BH mass estimators]{Shen_Liu2012}. For the targets in which only the CIV line fit is provided, we adopted the \cite{vestergaard2006} method.
%If two targets are analysed with the same method in the two works, we consider the $\lambda_{\rm Edd}$ value to which it is associated the smallest uncertainty. 
The resulting `observed' PDF of the $\lambda_{\rm Edd}$ values is shown in the left panel of Fig. \ref{lamda_sub-edd_comp_MV_Farina_S11}, indicated with the orange shaded region.\footnote{\citetalias{mazzucchelli2023} used bolometric correction by \cite{richards2006} to calculate $L_{\mathrm{bol}}$. These bolometric corrections were found to overestimate $L_{\mathrm{bol}}$ in the case of highly luminous quasars \citep{trakhtenbrot2012}. Thus, the actual PDF of $\lambda_{\rm Edd}$ may be shifted towards lower values.} 

%We have combined both the samples and in the case of common AGN we have considered value of $\lambda_{\rm Edd}$ with the least absolute uncertainties. 
%As adopted in \citetalias{farina2022}, we consider Eddington ratio calculated using \cite{shen2011} mass estimator from \citetalias{farina2022}. For the objects that only have a CIV line fit in \citetalias{farina2022}, we use Eddington ratio from \cite{vestergaard2006} mass estimator. For all the objects in \citetalias{mazzucchelli2023}, we use eddington ratio using \cite{shen2011} mass estimator. 

To compute $\lambda_{\rm Edd}$ values from \citetalias{Valentini:2021} simulation, we considered the accretion rate of the SMBHs that have masses consistent with the literature sample (namely $M_{\rm BH}>10^8$~M$_{\rm \odot}$) in the redshift range $6<z<7.5$. We ended up with two SMBHs that satisfy these criteria. We finally selected the SMBH with $ L_{\rm bol} >  2.7 \times 10^{46} \ \mathrm{L}_{\odot}$. This choice enables a proper comparison between observations and simulation since the observed sample consists of very luminous quasars (the least luminous source has $L_{\rm bol} = 2.86^{+0.17}_{-0.15} \times 10^{46} \ \mathrm{L}_{\odot}$). We evaluated  $\lambda_{\rm Edd}$ for the selected BH\footnote{This is the BH located at the centre of the most massive sub-halo. Further properties of this BH have been thoroughly analysed in \citetalias{Valentini:2021}.} at each time step of the simulation in the redshift range $6<z<7.5$. We considered each accretion episode as independent and computed the corresponding PDF. The resulting `simulated' PDF is shown in the left panel of Fig. \ref{lamda_sub-edd_comp_MV_Farina_S11} with a blue shaded region.

From the comparison between the observed and simulated PDF, it resulted that the median of these distributions (0.46 for the observed PDF and 0.44 for the simulated one) are perfectly consistent with each other (see the vertical orange and blue lines in the left panel of Fig. \ref{lamda_sub-edd_comp_MV_Farina_S11}). Further, we applied the two-sample Kolmogorov–Smirnov (K-S) test  \citep{kolmogorov1933sulla,smirnov1948table} to the observed and simulated PDFs.
%In fact, by applying the Kolmogorov–Smirnov (K-S) test  \citep{kolmogorov1933sulla,smirnov1948table} to the {\bf observed and simulated} PDFs, we find a $p$-value equal to 1.9, meaning that we can reject the null hypothesis that the two distributions are drawn from the same sample at less than 2-$\sigma$ (vertical blue line in the right panel of Fig. \ref{lamda_sub-edd_comp_MV_Farina_S11}).  
We accounted for the uncertainties in the observed PDF by adopting the bootstrap re-sampling technique, as done in \citetalias{farina2022}. To this aim, we first randomly re-sampled the observed PDF $N_{\rm re-samp}$ times, taking into account the errors associated to the observational values. Then, we included the average error of the literature sample ($\sim 0.1 \ \mathrm{dex}$, see Table 4 in \citetalias{farina2022} and Table 1 in \citetalias{mazzucchelli2023}) in the simulated PDF, and we randomly re-sampled it, as done for the observed one. Finally, for each $N_{\rm re-samp}$ couple, we performed the two-sample K-S test, and we associated the corresponding $\sigma$ values %\footnote{With the term $\sigma$ value we mean the following: if, for example, the p-value of the K-S test implies that the two PDFs are drawn from the same underlying distribution at the 2-$\sigma$ level, the corresponding $\sigma$ value is equal to 2.}. 
to the resulting p-values. 
%
%The p-values gives the probability that the simulated distribution reproduces the observed one. 
%
%P-value gives the probability that two cumulative distributions would be as far apart as observed given that both are randomly sampled from the identical distribution.} 

The right panel of Fig. \ref{lamda_sub-edd_comp_MV_Farina_S11} shows the PDF of the $\sigma$ values resulting from the bootstrap re-sampling technique considering $N_{\rm re-samp} = 10^6$. The median of the $\sigma$ value PDF is 1.8. %We find that the probability of rejecting the null hypothesis that these two samples are drawn from the same underlying distribution is not significant ($\sim$1\%), meaning that the simulated PDF provides a good representation of the observed one.
We found that the probability of having $\sigma$ values greater than three is very small ($\sim$1\%), implying that the probability of rejecting the null hypothesis that these two samples are drawn from the same underlying distribution is not significant. We thus concluded that the simulated PDF provides a good representation of the observed one.

In the \citetalias{Valentini:2021} simulation, however, the BH accretion is capped at the Eddington limit. To check whether and how much this assumption in the simulation affects the results, we used Eq. \ref{eq-Mdot-Bondi} to re-compute the accretion rate of the BHs for which $\lambda_{\rm Edd}=1$, and we repeated the K-S test analysis as detailed above. The resulting $\lambda_{\rm Edd}$ distribution from post processing is shown as a green shaded region in the left panel of Fig. \ref{lamda_sub-edd_comp_MV_Farina_S11}; the K-S test results are shown in the right panel of the same figure with a green shaded region. The new median of the $\sigma$ value PDF is 1.6 (instead of 1.8). The probability that we can reject the null hypothesis in this case is also $\sim$1\%. This simple test suggests that our conclusions are not biased by the Eddington cap prescription on the accretion. We further discuss this caveat in Sec. \ref{sec:discussion}.

To further validate our model, we compared the results of our simulation with the NIRSpec/PRISM sample by \cite{greene2023}, %\cite{2023arXiv230905714G} presented NIRSpec/PRISM spectroscopic follow-up of compact sources selected in the UNCOVER+Abell 2744 field. We had chosen AGNs from \cite{2023arXiv230905714G} samples 
which contains seven AGN with $L_{\rm bol} > 10^{44.2} \rm erg/s$ within the redshift range $5.8<z<8.5$. For these sources, the estimated Eddington ratios range from 0.04 to 0.4. Such a range is consistent with the mean $\lambda_{\rm Edd}$ value (0.14) found in our simulation by applying the same luminosity cut in the redshift range $ 6 < z < 8.5$.
Finally, we note that recent JWST spectroscopic observations of high-z (7<z<10) AGN \citep{bosman2023,furtak2023,kokorev2023} are consistent with sub-Eddington accretion rates ($\lambda_{\rm EDD}\sim 0.3-0.4$), with the exception of CEERS\textunderscore 1019, whose accretion rate is consistent with the Eddington limit \citep[$\lambda_{\rm EDD}=1.2\pm 0.5$;][]{larson2023}.

%of our simulation and compared their mean Eddington ratio ($0.21\pm0.28$) with mean Eddington ratio (0.15) of the luminosity controlled sample in our simulation. \cite{2023arXiv230905714G} has the lowest luminosity of $10^{44.2} \rm erg/s$ within the considered red-shift range therfore we have chosen all the AGNs with $L_{\rm bol} > 10^{44.2} \rm erg/s$ in our simulation for the comparision. 

%\mb{(Comment MB : Should we show the following comparison in table 1 between the properties of the host galaxy of chosen SMBH at z=10 and GNz11. We do not have snapshot for $z > 10$ so we can not show comparision at any other red shift closer to GNz11).}

\begin{center}
\begin{table}
\caption{Properties of the host galaxy of the most MBH in the simulation at z=10 and GN-z11 (from \cite{tacchella2023}).}
\begin{tabular}{c c c} 
 \hline
  & SIM value & Observational value \\ [0.5ex] 
 \hline
 log $M_{*}[\mathrm{M}_\odot]$ & 8.9 & 9.1$^{+0.3}_{-0.4}$ \\ 
 SFR $[\mathrm{M}_\odot~\mathrm{yr}^{-1}]$ & 5.4 & 21$^{+20}_{-10}$ \\ 
 %log $M_{\rm BH} [\mathrm{M}_\odot]$ & $6.5$ & $6.2$ \\ [1ex]
 \hline
\end{tabular}
\label{table:properties_host_galaxy_&_Gnz11}
\end{table}
\end{center}
%\ref{table:properties_host_galaxy_&_Gnz11} shows the comparision of stellar mass and SFR of the host galaxy of the most massive black hole in \citetalias{Valentini:2021} with the GNz11 properties. Stellar mass, $M_{*}$, from \citetalias{Valentini:2021} is within $1 \ \sigma$ and SFR is about $1.5 \ \sigma$ from the observational estimation presented in \cite{tacchella2023}. 

\section{Eddington ratio predictions against $z\sim 10$ data}
\label{sec:EVS_theory}
%Once that we have checked 
After confirming that the \citetalias{Valentini:2021} model is able to reproduce the accretion properties of the well-studied population of $z\sim 6$ quasars, we searched in the simulation for BHs with $M_{\rm BH}\sim 10^6 \ \mathrm{M}_\odot$ in the redshift range $10<z<11$ in order to investigate whether the same model can reproduce the properties of GN-z11 at $z=10.6$. We found one BH-galaxy system in our simulated volume that can be representative of GN-z11. The BH seeded at $z\sim 16.1$ has a mass of $=1.4\times 10^6 \ \mathrm{M}_\odot$ at $z=10.6$, which is fairly consistent with the BH mass of GN-z11; however, it is accreting at $\lambda_{\rm Edd}=0.6$, smaller than what is suggested for GN-z11. We further report in Table \ref{table:properties_host_galaxy_&_Gnz11} the properties of its host galaxy at $z=10$, namely at the closest snapshot in our simulation to the GN-z11 redshift. The galaxy properties are consistent (within 1.6-$\sigma$) with those of GN-z11 in terms of both stellar mass ($M_*$) and SFR, as computed by \citet{tacchella2023}. 

Furthermore, we computed the $\lambda_{\rm Edd}$ distribution of our GN-z11-like AGN with the same procedure adopted in Sec. \ref{sec:comparison}, and we report the results in Fig. \ref{fig:EVS_z10-11} with a shaded blue region. It is still possible that GN-z11 was detected while it was experiencing a rare episode of super-Eddington accretion. To compute the likelihood of this scenario, we relied on the EVS.
\subsection{The extreme value statistics}\label{EVS}
We considered a cumulative distribution function (CDF), $F(x)$, and drew a sequence of N random variates $\left\{X_{i}\right\}$ from it. We called $X_{max} \equiv {\rm sup}\left\{ X_{1} ... X_{N} \right\}$ the largest value of this sequence. If all variables are identically distributed and mutually independent,\footnote{We underscore that the hypothesis of independent accretion episodes is not satisfied in our approach since we are using the results obtained from a single zoom-in simulation. In Sec. \ref{sec:discussion}, we further discuss this point and how we plan to overcome this limitation in future works.} then the probability that all of the deviates are less than or equal to some $x$ is given by 
\begin{equation}  
    \label{eq-EVS_PDF_lessthan_equal} 
    \centering
    \begin{array}{cl}
    \Phi\left(X_{\rm max} \le x; N \right) & = F_{1}\left( X_{1} \le x\right) .. F_{N}\left( X_{N} \le x\right) \\ 
    & = F^{N} (X). \\
    \end{array}
\end{equation}
The PDF of $X_{\rm max}$ can then be obtained by differentiating Eq. \ref{eq-EVS_PDF_lessthan_equal} with respect to $x$: 
\begin{equation}  
\label{eq-EVS_PDF} 
\centering
    \begin{array}{cl}
    \Phi\left(X_{\rm max} = x; N \right) & = N \ F'(x) \left[ F(x)\right]^{N-1} \\
    & = N \ f(x) \left[ F(x)\right]^{N-1},
    \end{array}
\end{equation}
where $f(x) = dF(x)/dx$ is the PDF of the considered distribution. Equation \ref{eq-EVS_PDF} provides the probability of finding the extreme value $X_{\rm max}$ after randomly extracting $N$ variables from a given distribution $f_{x}$.

%PDF of the re-calculated Eddington ratio (in the postprocessing using Eq. \ref{eq-Mdot-Bondi}) for the accretion episodes satisfying the above BH mass and redshift criteria is shown in the blue bars in Figure \ref{fig:EVS_z10-11}.
%Now to predict the probability of super-Eddington accretion rate from our simulations, we have consider BH with $M_{\rm BH}>10^6 \mathrm{M}_\odot$ in our simulation within $10<z<11$ similar to the GN-z11 BH. 

\subsection{Application of the extreme value statistics to GN-z11}

\begin{figure}
\begin{center}
\includegraphics[width=0.45\textwidth]{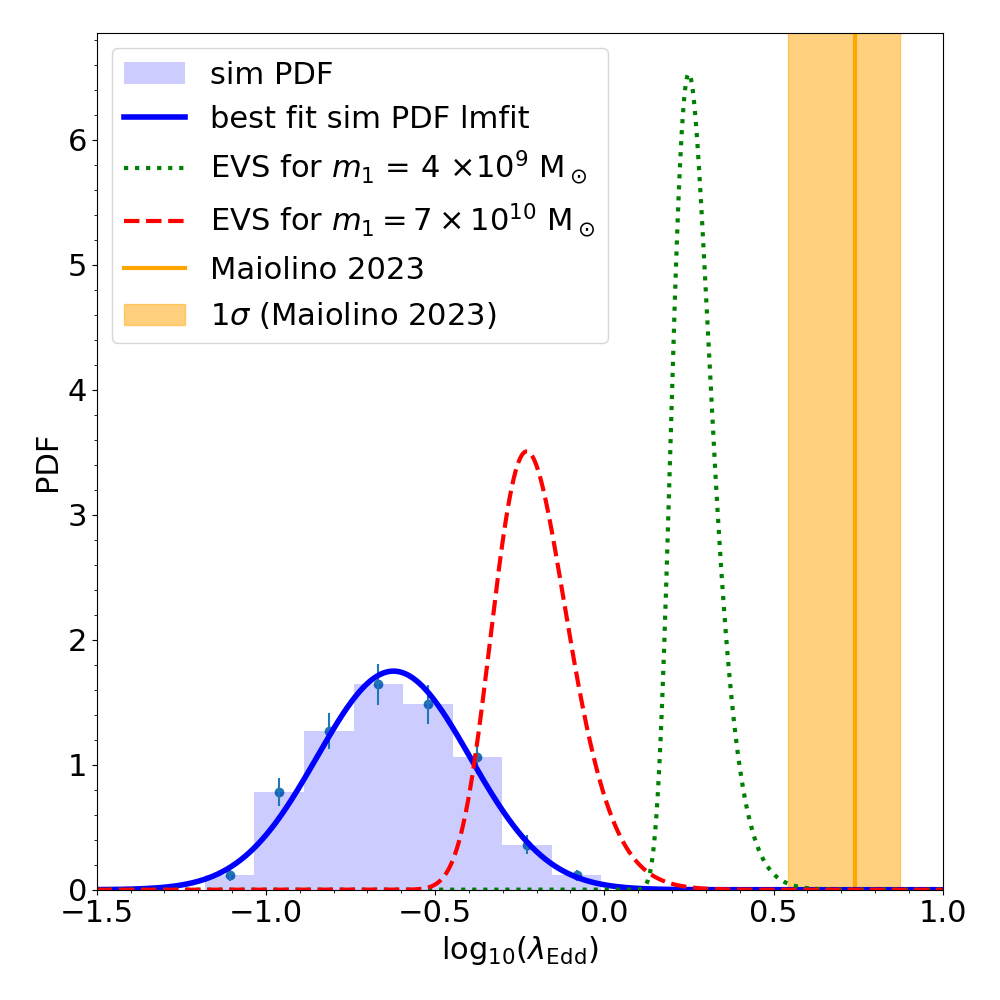}
\caption{Extreme value statistics PDF as a function of the Eddington ratio for a GN-z11-like AGN. Blue solid line shows the best-fit Gaussian (reduced $\chi^2_{\nu}=1.4$) to the Eddington ratio distribution from the \citetalias{Valentini:2021} simulation for BHs in the redshift range $10<z<11$ and $M_{\rm BH} > 10^6 M_{\odot}$. The peak of the best fit is around $\lambda_{\rm Edd} \sim 0.25$. The dashed (dotted) red (green) line represents the EVS for the largest (smallest) considered value of $m_1$ (see text). The orange solid line (shaded region) shows the GN-z11 Eddington ratio (1$\sigma$ uncertainty) estimated by \cite{maiolino2023a}.} %Please note that for "$N=1$" (yellow dashed line) EVS PDF is the same as PDF of random variable.}
\label{fig:EVS_z10-11}
\end{center}
\end{figure}

To apply the EVS to the case of GN-z11, we needed to know both (i) the functional form describing the PDF and CDF ($f(x)$ and $F(x)$, respectively, in Eq. \ref{eq-EVS_PDF}) of the simulated $\lambda_{\rm Edd}$ distribution and (ii) the number of random extractions ($N$ in Eq. \ref{eq-EVS_PDF}) that apply to the case of JWST observations. 
For what concerns (i), we simply fit the blue shaded region in Fig. \ref{fig:EVS_z10-11} with a Gaussian (solid blue line in Fig. \ref{fig:EVS_z10-11}), and we computed the corresponding CDF\footnote{We adopt the python script \href{https://numpy.org/doc/stable/reference/generated/numpy.cumsum.html}{{np.cumsum}}.}. When fitting the simulated PDF with a Gaussian (solid blue line), the tail extends beyond the Eddington cap. %thus providing a non-vanishig probability of having values $\lambda > \lambda_{\rm Edd}$ .
Thus, even if the \citetalias{Valentini:2021} simulations are capped at the Eddington limit, the probability of having extreme events, such as super-Eddington accretion episodes, is non-vanishing.

The calculation of (ii) was less trivial. We associated the number $N$ of random extractions to the number of DM halos contained in the volume\footnote{A more recent estimate provides a smaller volume \citep[$2.2\times 10^5 \rm Mpc^3$;][]{naidu2022} than the one adopted here. A smaller volume would imply a smaller $N$ value, thus strengthening the main result of our work.} $V$ ($\sim 1.2 \times 10^6 \ \mathrm{Mpc}^3$) covered by the observations \citep{oesch2016} that discovered GN-z11 in the CANDELS field. We thus adopted the following relation:
\begin{equation}  
\label{eq-EVS_N_GNz11} 
N = V \int_{m1}^{m2}  \ dM \ \left( \frac{dn(M)}{dM} \right)_{z=10.6},
\end{equation}
where $m_{1}$ and $m_{2}$ represent the minimum and maximum mass of the DM halo that are expected to host GN-z11 and $\frac{dn(M)}{dM}$ is the halo mass function, for which we considered the \cite{sheth1999} functional form. We assumed $m_2=10^{13}~\rm M_{\odot}$ because we noticed that larger values do not change our results since the integral already converges with the assumed $m_2$ value. To estimate $m_{1}$, we adopted three different and independent approaches that rely on the SFR, $M_*$, and $M_{\rm BH}$ estimates of GN-z11 (see Appendix \ref{halognz11}). We derived that GN-z11 is hosted by a DM halo with a fiducial mass\footnote{This estimate is consistent with the results by \cite{scholtz2023} and \cite{tacchella2023}, according to which $M_{\rm h}\sim3-8\times 10^{10} \ ~\rm M_{\odot}$.} of $M_{\rm h}\sim 0.7-4\times 10^{11} \ \mathrm{M}_\odot$. However, values as small as $M_{\rm h}\sim 4\times 10^{9}~\rm M_{\odot}$ are not excluded by our analysis. We found $N=0.006$, for $m_1 = 4 \times 10^{11} \ \mathrm{M}_\odot$ %largest value of $m_1$
, that is, no DM halos of this mass are expected to be in the survey volume. Hence, we considered $7 \times 10^{10} \ \mathrm{M}_{\odot}$ as the largest value for $m_{1}$ and $4 \times 10^{9} \ \mathrm{M}_{\odot}$ as the smallest.

Figure. \ref{fig:EVS_z10-11} shows the EVS for the different values of $m_1$ mentioned above (see also Table \ref{table:EVS_Probs}). For the smallest value of $m_1$, namely the one providing the largest number of DM halos ($N=16416$), $P\left( \lambda_{\rm Edd} > 1\right) = 0.98$ and $P\left( \lambda_{\rm Edd} > 3.5\right) = 2 \times 10^{-3}$. This means that our model allows for a non-vanishing probability of  super-Eddington accretion episodes if $m_1=4\times 10^{9} \ \mathrm{M}_\odot$, but the probability of having $\lambda_{\rm Edd}$ values as large as the one observed in GN-z11 is very small in any case.

We repeated the same calculation of the accretion rate for the few ($\sim2\%$) $\lambda_{\rm Edd}=1$ episodes by removing the Eddington cap. We found that the results reported in Table \ref{table:EVS_Probs} do not change. The maximum value of the Eddington ratio obtained in this case is $\lambda_{\rm Edd}=2.24$. Such a value is smaller than that reported in \citet{maiolino2023a}. 

%{\bf Additionally, we have also applied different bolometric cutoffs (same as done in the sec. \ref{sec:comparison}) to quantify its impact. As applying cutoff will only chose the large values, the resulting Gaussian fit has higher mean but lower standard deviation. Thus the final probability calculated in this work does not change significantly.} 

In summary, we quantified the probability that a GN-z11-like MBH, though accreting on average at $\lambda_{\rm EDD}=0.25$, was detected while experiencing a rare episode of super-Eddington accretion. We find that this probability is very small ($< 2\times 10^{-3}$).

\begin{table}
\begin{center}
%\caption{EVS probability for $\lambda > (5.5-1 \sigma = 3.5) $ and $\lambda > (5.5-2 \sigma = 1.5) $  for the estimated $\lambda = 5.5 \pm 2 $ of GN-z11 black hole (\cite{maiolino2023a})}
\caption{Extreme value statistics probability for $\lambda_{\rm Edd} > 1$ and $>3.5$ (namely the 1$\sigma$ lower limit of the estimate by \citet{maiolino2023a} for GN-z11) resulting from different values of $m_{1}$. (See Eq. \ref{eq-EVS_N_GNz11}).}
\begin{tabular}{c c c c} 
%\vspace{0.1cm}
 \hline
 $m_{1}$[$\mathrm{M}_{\odot}$] & $N$ & P($\lambda_{\rm Edd}$>1) & P($\lambda_{\rm Edd}$>3.5)  \\ [0.5ex] 
 \hline
%$  5 \times 10^9 $ & 11334  & 0.99 & $3 \times 10^{-3}$ \\
$  4 \times 10^9 $ & 16416  & 0.98 & $ 2 \times 10^{-3}$ \\
% 882 & \ $   1.6 \ \times 10^{10} $ & 0.22 & $2 \times 10^{-4}$ \\
 $   7 \times 10^{10} $ & 20 & 0.06 & $ 3\times 10^{-6}$\\
 [1ex] 
 \hline
\end{tabular}
\label{table:EVS_Probs}
\end{center}
\end{table}

%\subsection{Prediction for future surveys}

%\cite{2019A&A...631A..85E} is expected to find $~100$ quasars at $7<z<8$.
%To calculate the probable range of the DM halo mass for GNz11, we consider the observational estimation of stellar mass derived from NIRCam photometry of $log \left( \frac{M_{*}}{M}  = 9.1^{+0.3}_{-0.4}\right)$ presented in \cite{2023ApJ...952...74T}. Considering the star formation efficiency and the cosmic baryon to dark matter fraction, we can estimate the DM halo mass as follows,

%\begin{equation}  
%\label{eq-EVS_M_DM} 
%M_{DM} = \frac{M_{*}}{\eta_{SF}} \ \left( \frac{\Omega_{DM}}{\Omega_{b}} \right)  
%\end{equation}
%where $\eta_{SF}$ is the star formation efficiency, $\Omega_{b}$ is the baryon density and $\Omega_{DM}$ is DM density. We have taken \cite{2020A&A...641A...6P} values for $\Omega_{b} = 0.02$ and $\Omega_{DM} = 0.12$. We have used stellar mass range within $3 \sigma$ level : $ 7.9 < log \left( \frac{M_{*}}{M}  \right) <10.0$ 
%(from \cite{2023ApJ...952...74T}). To get the most conservative range of $M_{DM}$, we use lower limit of stellar mass with large value of star formation efficiency to get lower bound, and upper limit of stellar mass with less star formation efficiency to get upper bound. Putting all the information in Equation \ref{eq-EVS_M_DM}, the DM halo mass range will be,

%\begin{equation}  
%\label{eq-EVS_M_DM_range} 
%log \left( \frac{M_{DM}}{M_\odot} \right) %< 
%end{equation}
\section{Summary and discussion}
\label{sec:discussion}
In this work, we have investigated the probability of having an MBH ($\sim 10^6 \ \mathrm{M_{\rm \odot}}$) at $z\sim 10-11$ accreting at the super-Eddington rate ($\lambda_{\rm Edd}\sim 5.5$) similar to the one recently detected by JWST in GN-z11 \citep{maiolino2023a}.
We first considered the accretion properties of simulated accreting SMBHs at $z\sim 6-7.5$ as provided by the zoom-in simulations developed by \citet{Valentini:2021}. By comparing the simulated $\lambda_{\rm Edd}$ PDF with the results from a state-of-the-art sample of $z\sim6$ quasars \citep{farina2022,mazzucchelli2023}, we found that our accretion model successfully reproduces the data.
% By comparing the simulated $\lambda_{\rm Edd}$ probability distribution function (PDF) with the results from the XQR-30 sample \citep{dodorico2023} of $z\sim 6$ quasars \citep{farina2022,mazzucchelli2023} we find that our accretion model successfully reproduces the data. 

We then analysed the $\lambda_{\rm Edd}$ PDF of MBHs at $z\sim 10-11$ and computed the EVS. Assuming that the observations that discovered GN-z11 cover a volume of $\sim 1.2 \times 10^6 \ \mathrm{Mpc}^3$ \citep{oesch2016}, we find that the probability of finding a GN-z11-like object is very small ($\sim 10^{-3}$). Our result does not exclude the presence of an accreting MBH that is partly powering the GN-z11 luminosity. However, the BH-accretion and AGN-feedback recipes adopted in our model cannot reproduce the total observed luminosity of GN-z11. %If the observed emission line fluxes have not been over-estimated, 
The discrepancy with JWST data suggests that the missing physics in our model (detailed below) have to boost the luminosities of $z\sim 10$ MBHs to the observed fluxes while still reproducing the $\lambda_{\rm EDD}$ PDF of $z\sim 6$ quasars.

%In this \textit{Letter}, we have derived the simulated Eddington ratio distribution for BH in the range $6<z<7.5$ and compared it with that observed in the same redshift range by \citetalias{farina2022}. Having established the robustness of our simulated results, we have applied Extreme Value Statistics (EVS) to calculate the EVS PDF of the Eddington ratio for GN-z11-like MBH with mass $M_{\rm BH} > 10^6 \mathrm{M}_\odot$ in the redshift range $10<z<11$. Our main conclusions are as follows:
%The predicted probability of finding MBH in the considered mass and redshift ranges with accretion rates $>3.5$ times the Eddington one is $3 \times  10^{-3}$. 

%    \item \af{This item is rather idle... what is the point here?} \mb{I was thinking of making point that these probabilities can also be useful for future MBH in 10<z<11. But may be we can add this at end of discussion.} The probabilities of finding sources accreting at super-Eddington rates is given in  Table \ref{table:EVS_Probs} are valid for MBH in the redshift range of $10 < z < 11$ and $M_{\rm BH} > 10^6  \mathrm{M}_\odot$ for the survey size $\sim10^6 \ \mathrm{Mpc}$.
%\end{itemize}
% Finally, we have given the EVS prescription for redshift range $6<z<11$ for future surveys.

One limitation of our work is that we are interpreting the super-Eddington accretion of GN-z11 with a model that caps the accretion to the Eddington limit. %{\bf We justify this approximation with the fact that the \citetalias{Valentini:2021} accretion model has very small fraction of accretion episodes ($\sim 2\%$) which are affected by the Eddington cap at $\lambda_{Edd}= 1$} (see the blue distribution in Fig. \ref{fig:EVS_z10-11}). %that the probability of having accretion episodes at $\lambda_{Edd}= 1$ values is extremely small ($\sim 2\%$)
%We justify this approximation with the fact that the \citetalias{Valentini:2021} accretion model predicts that the probability of having accretion episodes at $\lambda_{Edd}= 1$ is extremely small ($\sim 2\%$). 
We justify this approximation with the fact that the \citetalias{Valentini:2021} accretion model predicts that the probability of having maximal accretion ($\lambda_{Edd}= 1$) is extremely small ($\sim 2\%$; see the blue distribution in Fig. \ref{fig:EVS_z10-11}). Thus, the assumption of the Eddington cap affects only a negligible number of accretion episodes. It is not straightforward to quantify the impact of the Eddington cap on the actual accretion rate. This is because feedback from super-Eddington episodes can greatly affect the surrounding medium,  possibly quenching subsequent accretion events. Such an expectation is corroborated by the results in \citet[][see Fig. 11 and 12]{massonneau2023}. These authors compared the accretion rate evolution of MBHs as predicted by simulations with and without super-Eddington accretion and feedback. They found that models that account for super-Eddington accretion result in a smaller fraction of Eddington accretion episodes compared to the Eddington-capped case. This means that if the Eddington cap is removed in our simulation, the $\lambda_{Edd}$ PDF would shift towards smaller values, thus yielding an even smaller EVS probability for $\lambda_{Edd}>1$.
If we consider the results from other studies allowing super-Eddington accretion rates onto BHs, the occurrence of $\lambda_{\rm Edd} \sim 5.5$ events is unlikely. To start with, in \cite{massonneau2023}, super-Eddington accretion events reach peak values of only two to three times the Eddington limit. %Moreover, after each super-Eddington events, the accretion rapidly drops to the sub-Eddington level. The fraction of time that the BH spent close to the Eddington limit becomes much smaller in the super-Eddington models compared to the Eddington limited case. In our work, when we consider super-Eddington accretion in post process, we are neglecting this effect since we are not taking into account super-Eddington feedback. The latter would shift the PDF of $\lambda_{\rm Edd}$ towards lower values thus the EVS prediction of our fiducial model would provide an upper limit to the actual probability.
%The limitation of our work is that the simulation does not include the super Eddington feedback. However other works \cite{massonneau2023} allowing super Eddington accretion suggested that super Eddington feedback has strong impact on the surrounding of the BH and thus after brief super Eddington phases, AGN spends significant fraction of remaining time in low accretion. It is also observed in their work that the feedback events are not impacting much on the surrounding of the BH in Eddington limited accretion case thus BH accretes closer to the Eddington limit and reaches self regularisation only near the end of the simulation. These results suggest that our Eddington limited accretion case can provide the upper limit on our EVS prediction, as allowing strong super Eddington feedback can possibly shift the Eddington ratio PDF towards the lower mean. 
 \cite{zhu2022} performed zoom-in hydrodynamical simulations of $z>6$ SMBHs, exploring models with different seeding prescriptions, radiative efficiencies, accretion, and feedback models. In these simulations, the most extreme accretion event for MBHs is characterised by $\lambda_{\rm Edd} = 2.6 $, and it only occurs at $z < 10$.  
Furthermore, in the radiation-hydrodynamic study by  \cite{pacucci2015}, the average Eddington ratio of the considered MBH is $\lambda_{\rm Edd} \sim 1.35$. The accretion halts when the emitted luminosity reaches the value $\sim 3 L_{\rm Edd}$. 

%{\bf \cite{schneider2023} found that GN-z11 can be powered by $\lambda_{\rm Edd} =~ 2-3 $ accreting BH considering super-Eddington BH accretion scenario using semi-analytical model.} 
Finally, even state-of-the-art simulations \citep[e.g.][but see also 
\citealt{jeon2024}]{lupi2023} that account for the entire range of accretion rates (from advection-dominated accretion flow to super-Eddington regimes) cannot explain the observational results of GN-z11. 
%They use physically motivated prescription to account the unresolved dynamical friction effects rather than moving BH at minimum of potential in each time step. 
According to these calculations, $10^6~\rm M_{\odot}$ BH can experience super-Eddington accretion events (up to $\lambda_{\rm Edd} \sim 10-100$) when the MBH settles in the centre of the galaxy and the potential well becomes deep enough to sustain ingent gas inflows. This phase can be reached only after the supernova feedback stops perturbing the gas efficiently enough to hamper BH accretion. In this scenario, super-Eddington events only occur at $z<9.5$, namely $\sim500~\rm Myr$ after the MBH seeding. %as at higher $z$ supernova feedback strongly perturbs the gas in galaxy thus preventing the MBH which is offset from the minimum potential to accrete efficiently.
% Super-Eddington accretion events (up to λEdd ∼ 10 − 100, in this case) onto a 106 M⊙ BH only occurs at z < 9.5 as before BH was not at the minimum of the potential and supernova feedback strongly perturbs the gas in galaxy preventing the efficient MBH accretion

%As the radiative feedback from the accretion is not considered in \cite{smith2018}, it provides an upper limit of accretion and overall growth of black holes. Eddington ratio distribution from our simulation, shown in the Fig. \ref{fig:EVS_z10-11} with blue bars, also has very small fraction of BHs accreting super-Eddington similar to \cite{smith2018}.   

Another caveat is that we have used results from a single zoom-in simulation. This is clearly insufficient to perform an extensive statistical analysis of the problem. In fact, when comparing the observed $\lambda_{\rm Edd}$ distribution with the simulation, we are considering a simulated MBH evolving from $10^8 \ \mathrm{M}_\odot$ at $z=7.5$ to $10^9 \ \mathrm{M}_\odot$ at $z=6$. In other words, we are making the strong assumption of associating different observed quasars to an accreting SMBH captured at different times during its accretion history. 

Another assumption in our model that may affect the final results of our work concerns the seeding recipe. We have in fact considered only the possibility that SMBH can originate from heavy seeds (seed mass $\sim10^5 \ \mathrm{M}_\odot$). It is not straightforward to foresee how this assumption might affect the simulated $\lambda_{\rm Edd}$ PDF and  corresponding EVS. We can, however, mention that models considering the light seed scenario and allowing for super-Eddington accretion are unlikely to explain the extreme $\lambda_{\rm Edd}$ value found in GN-z11. For example, \cite{smith2018} post processed the Renaissance simulation to follow the growth of BHs from individual remnants of Pop III stars, finding that (i) the accretion rate was very small ($ < 10^{-4} \ \lambda_{\rm Edd}$) for the majority of the $\sim15000$ BHs, (ii) only a small fraction $(2-3 \%)$ of BHs accreted with $\lambda_{\rm Edd} > 10^{-4}$, and (iii) only one BH from $\sim$15000 BHs in a comoving volume of $\left(40 \ \rm Mpc\right)^3$ accreted at the maximum accretion rate of $\lambda_{\rm Edd}\sim 3$, for a single timestep.

The robustness of the main conclusion of our work, namely the low probability of having a $10^6~\rm M_{\odot}$ super-Eddington accreting BH at $z\sim 10$, is weakened by the assumptions discussed above and the limitations of our approach. Still, our model proposes a novel method to link JWST observations with theoretical models in order to interpret, in particular, the overabundance of AGNs found at high-$z$ \citep[e.g.][]{maiolino2023b,harikane2023}.  

The most promising way to overcome the limitations of our work is to apply the EVS method to multiple large-scale simulations \citep[e.g.][]{habouzit2021,habouzit2022} that consider different seeding prescriptions and that also include light and intermediate seeds \citep{zhu2022}. Furthermore, for a fair comparison with observed super-Eddington accretion events, it is important to properly model super-Eddington accretion in simulations adopting the radiatively inefficient slim-disc solution \citep{sadowski2009}. In this case, in fact, there is not a linear relation between $\lambda_{\rm Edd}$ and $\dot{M}_{\rm BH}/\dot{M}_{\rm Edd}$; in particular, the measured value of $\lambda_{\rm Edd}$ is expected to plateau at $\sim 3$ for $\dot{M}_{\rm BH}/\dot{M}_{\rm Edd}>10$ \citep{madau2014}. 

From the observational point of view, further detections and characterisation of sources similar to GN-z11 in the early Universe will be fundamental to better constrain theoretical models and test their predictions against a larger data sample. In order to find rare sources, such as GN-z11, it is necessary to survey a P($\lambda_{\rm Edd}$>3.5)$^{-1}$ $\sim 500\times$ larger volume with photometric coverage at observed wavelengths greater than $ 1\mu$m and to then follow them up spectroscopically. 
%Moreover in the future, similar to our approach can be utilize on the hydro-dynamical simulation without Eddington limited accretion and feedback. 
 %this approach provides the upper limit of the EVS PDF as the super Eddington feedback reduced the gas in the surrounding and decrease the accretion to sub Eddington just after small episode of super Eddington accretion. 

\label{sec:conclusion}
\begin{acknowledgements} 
MB acknowledges support from PNRR funds. SG acknowledges support from the PRIN 2022 project (2022TKPB2P), titled: "BIG-z: Building the Giants: accretion, feedback and assembly in z>6 quasars". MV is supported by the Fondazione ICSC National Recovery and Resilience Plan (PNRR), Project ID CN-00000013 "Italian Research Center on High-Performance Computing, Big Data and Quantum Computing" funded by MUR - Next Generation EU. MV also acknowledges partial financial support from the INFN Indark Grant. E.P.F. is supported by the international Gemini Observatory, a program of NSF’s NOIRLab, which is managed by the Association of Universities for Research in Astronomy (AURA) under a cooperative agreement with the National Science Foundation, on behalf of the Gemini partnership of Argentina, Brazil, Canada, Chile, the Republic of Korea, and the United States of America.
\end{acknowledgements}

\bibliographystyle{aa_url}
\bibliography{biblio}

\begin{thebibliography}{101}
\expandafter\ifx\csname natexlab\endcsname\relax\def\natexlab#1{#1}\fi

\bibitem[{{Ba{\~n}ados} {et~al.}(2018){Ba{\~n}ados}, {Venemans}, {Mazzucchelli}, {Farina}, {Walter}, {Wang}, {Decarli}, {Stern}, {Fan}, {Davies}, {Hennawi}, {Simcoe}, {Turner}, {Rix}, {Yang}, {Kelson}, {Rudie}, \& {Winters}}]{banados2018}
{Ba{\~n}ados}, E., {Venemans}, B.~P., {Mazzucchelli}, C., {et~al.} 2018, \href{http://dx.doi.org/10.1038/nature25180}{\color{magenta}\nat}, \href{https://ui.adsabs.harvard.edu/abs/2018Natur.553..473B}{553, 473}

\bibitem[{{Barkana} \& {Loeb}(2001)}]{barkana2001}
{Barkana}, R. \& {Loeb}, A. 2001, \href{http://dx.doi.org/10.1016/S0370-1573(01)00019-9}{\color{magenta}\physrep}, \href{https://ui.adsabs.harvard.edu/abs/2001PhR...349..125B}{349, 125}

\bibitem[{{Bondi}(1952)}]{bondi1952}
{Bondi}, H. 1952, \href{http://dx.doi.org/10.1093/mnras/112.2.195}{\color{magenta}\mnras}, \href{https://ui.adsabs.harvard.edu/abs/1952MNRAS.112..195B}{112, 195}

\bibitem[{{Bondi} \& {Hoyle}(1944)}]{bondi1944}
{Bondi}, H. \& {Hoyle}, F. 1944, \href{http://dx.doi.org/10.1093/mnras/104.5.273}{\color{magenta}\mnras}, \href{https://ui.adsabs.harvard.edu/abs/1944MNRAS.104..273B}{104, 273}

\bibitem[{{Bosman} {et~al.}(2023){Bosman}, {{\'A}lvarez-M{\'a}rquez}, {Colina}, {Walter}, {Alonso-Herrero}, {Ward}, {{\"O}stlin}, {Greve}, {Wright}, {Bik}, {Boogaard}, {Caputi}, {Costantin}, {Eckart}, {Garc{\'\i}a-Mar{\'\i}n}, {Gillman}, {G{\"u}del}, {Henning}, {Hjorth}, {Iani}, {Ilbert}, {Jermann}, {Labiano}, {Lagage}, {Langeroodi}, {Pei{\ss}ker}, {Ray}, {Rinaldi}, {Topinka}, {van Dishoeck}, {van der Werf}, \& {Vandenbussche}}]{bosman2023}
{Bosman}, S. E.~I., {{\'A}lvarez-M{\'a}rquez}, J., {Colina}, L., {et~al.} 2023, \href{https://ui.adsabs.harvard.edu/abs/2023arXiv230714414B}{\href{http://dx.doi.org/10.48550/arXiv.2307.14414}{\color{magenta}arXiv e-prints}, arXiv:2307.14414}

\bibitem[{{Bouwens} {et~al.}(2010){Bouwens}, {Illingworth}, {Gonz{\'a}lez}, {Labb{\'e}}, {Franx}, {Conselice}, {Blakeslee}, {van Dokkum}, {Holden}, {Magee}, {Marchesini}, \& {Zheng}}]{bouwens2010}
{Bouwens}, R.~J., {Illingworth}, G.~D., {Gonz{\'a}lez}, V., {et~al.} 2010, \href{http://dx.doi.org/10.1088/0004-637X/725/2/1587}{\color{magenta}\apj}, \href{https://ui.adsabs.harvard.edu/abs/2010ApJ...725.1587B}{725, 1587}

\bibitem[{{Bunker} {et~al.}(2023){Bunker}, {Saxena}, {Cameron}, {Willott}, {Curtis-Lake}, {Jakobsen}, {Carniani}, {Smit}, {Maiolino}, {Witstok}, {Curti}, {D'Eugenio}, {Jones}, {Ferruit}, {Arribas}, {Charlot}, {Chevallard}, {Giardino}, {de Graaff}, {Looser}, {L{\"u}tzgendorf}, {Maseda}, {Rawle}, {Rix}, {Del Pino}, {Alberts}, {Egami}, {Eisenstein}, {Endsley}, {Hainline}, {Hausen}, {Johnson}, {Rieke}, {Rieke}, {Robertson}, {Shivaei}, {Stark}, {Sun}, {Tacchella}, {Tang}, {Williams}, {Willmer}, {Baker}, {Baum}, {Bhatawdekar}, {Bowler}, {Boyett}, {Chen}, {Circosta}, {Helton}, {Ji}, {Kumari}, {Lyu}, {Nelson}, {Parlanti}, {Perna}, {Sandles}, {Scholtz}, {Suess}, {Topping}, {{\"U}bler}, {Wallace}, \& {Whitler}}]{bunker2023}
{Bunker}, A.~J., {Saxena}, A., {Cameron}, A.~J., {et~al.} 2023, \href{http://dx.doi.org/10.1051/0004-6361/202346159}{\color{magenta}\aap}, \href{https://ui.adsabs.harvard.edu/abs/2023A&A...677A..88B}{677, A88}

\bibitem[{{Chongchitnan} \& {Silk}(2012)}]{chongchitnan2012}
{Chongchitnan}, S. \& {Silk}, J. 2012, \href{http://dx.doi.org/10.1103/PhysRevD.85.063508}{\color{magenta}\prd}, \href{https://ui.adsabs.harvard.edu/abs/2012PhRvD..85f3508C}{85, 063508}

\bibitem[{{Chongchitnan} \& {Silk}(2021)}]{chongchitnan2021}
{Chongchitnan}, S. \& {Silk}, J. 2021, \href{http://dx.doi.org/10.1103/PhysRevD.104.083018}{\color{magenta}\prd}, \href{https://ui.adsabs.harvard.edu/abs/2021PhRvD.104h3018C}{104, 083018}

\bibitem[{{Coatman} {et~al.}(2017){Coatman}, {Hewett}, {Banerji}, {Richards}, {Hennawi}, \& {Prochaska}}]{coatman2017}
{Coatman}, L., {Hewett}, P.~C., {Banerji}, M., {et~al.} 2017, \href{http://dx.doi.org/10.1093/mnras/stw2797}{\color{magenta}\mnras}, \href{https://ui.adsabs.harvard.edu/abs/2017MNRAS.465.2120C}{465, 2120}

\bibitem[{{Colombi} {et~al.}(2011){Colombi}, {Davis}, {Devriendt}, {Prunet}, \& {Silk}}]{colombi2011}
{Colombi}, S., {Davis}, O., {Devriendt}, J., {Prunet}, S., \& {Silk}, J. 2011, \href{http://dx.doi.org/10.1111/j.1365-2966.2011.18563.x}{\color{magenta}\mnras}, \href{https://ui.adsabs.harvard.edu/abs/2011MNRAS.414.2436C}{414, 2436}

\bibitem[{{Davis} {et~al.}(2011){Davis}, {Devriendt}, {Colombi}, {Silk}, \& {Pichon}}]{davis2011}
{Davis}, O., {Devriendt}, J., {Colombi}, S., {Silk}, J., \& {Pichon}, C. 2011, \href{http://dx.doi.org/10.1111/j.1365-2966.2011.18286.x}{\color{magenta}\mnras}, \href{https://ui.adsabs.harvard.edu/abs/2011MNRAS.413.2087D}{413, 2087}

\bibitem[{{Dayal} {et~al.}(2014){Dayal}, {Ferrara}, {Dunlop}, \& {Pacucci}}]{dayal2014}
{Dayal}, P., {Ferrara}, A., {Dunlop}, J.~S., \& {Pacucci}, F. 2014, \href{http://dx.doi.org/10.1093/mnras/stu1848}{\color{magenta}\mnras}, \href{https://ui.adsabs.harvard.edu/abs/2014MNRAS.445.2545D}{445, 2545}

\bibitem[{{Decarli} {et~al.}(2022){Decarli}, {Pensabene}, {Venemans}, {Walter}, {Ba{\~n}ados}, {Bertoldi}, {Carilli}, {Cox}, {Fan}, {Farina}, {Ferkinhoff}, {Groves}, {Li}, {Mazzucchelli}, {Neri}, {Riechers}, {Uzgil}, {Wang}, {Wang}, {Weiss}, {Winters}, \& {Yang}}]{decarli2022}
{Decarli}, R., {Pensabene}, A., {Venemans}, B., {et~al.} 2022, \href{http://dx.doi.org/10.1051/0004-6361/202142871}{\color{magenta}\aap}, \href{https://ui.adsabs.harvard.edu/abs/2022A&A...662A..60D}{662, A60}

\bibitem[{{Decarli} {et~al.}(2018){Decarli}, {Walter}, {Venemans}, {Ba{\~n}ados}, {Bertoldi}, {Carilli}, {Fan}, {Farina}, {Mazzucchelli}, {Riechers}, {Rix}, {Strauss}, {Wang}, \& {Yang}}]{decarli2018}
{Decarli}, R., {Walter}, F., {Venemans}, B.~P., {et~al.} 2018, \href{http://dx.doi.org/10.3847/1538-4357/aaa5aa}{\color{magenta}\apj}, \href{https://ui.adsabs.harvard.edu/abs/2018ApJ...854...97D}{854, 97}

\bibitem[{{Devecchi} {et~al.}(2012){Devecchi}, {Volonteri}, {Rossi}, {Colpi}, \& {Portegies Zwart}}]{devecchi2012}
{Devecchi}, B., {Volonteri}, M., {Rossi}, E.~M., {Colpi}, M., \& {Portegies Zwart}, S. 2012, \href{http://dx.doi.org/10.1111/j.1365-2966.2012.20406.x}{\color{magenta}\mnras}, \href{https://ui.adsabs.harvard.edu/abs/2012MNRAS.421.1465D}{421, 1465}

\bibitem[{{D'Odorico} {et~al.}(2023){D'Odorico}, {Ba{\~n}ados}, {Becker}, {Bischetti}, {Bosman}, {Cupani}, {Davies}, {Farina}, {Ferrara}, {Feruglio}, {Mazzucchelli}, {Ryan-Weber}, {Schindler}, {Sodini}, {Venemans}, {Walter}, {Chen}, {Lai}, {Zhu}, {Bian}, {Campo}, {Carniani}, {Cristiani}, {Davies}, {Decarli}, {Drake}, {Eilers}, {Fan}, {Gaikwad}, {Gallerani}, {Greig}, {Haehnelt}, {Hennawi}, {Keating}, {Kulkarni}, {Mesinger}, {Meyer}, {Neeleman}, {Onoue}, {Pallottini}, {Qin}, {Rojas-Ruiz}, {Satyavolu}, {Sebastian}, {Tripodi}, {Wang}, {Wolfson}, {Yang}, \& {Zanchettin}}]{dodorico2023}
{D'Odorico}, V., {Ba{\~n}ados}, E., {Becker}, G.~D., {et~al.} 2023, \href{http://dx.doi.org/10.1093/mnras/stad1468}{\color{magenta}\mnras}, \href{https://ui.adsabs.harvard.edu/abs/2023MNRAS.523.1399D}{523, 1399}

\bibitem[{{Eddington}(1926)}]{eddington1926book}
{Eddington}, A.~S. 1926, {The Internal Constitution of the Stars}

\bibitem[{{Eisenstein} {et~al.}(2023){Eisenstein}, {Willott}, {Alberts}, {Arribas}, {Bonaventura}, {Bunker}, {Cameron}, {Carniani}, {Charlot}, {Curtis-Lake}, {D'Eugenio}, {Endsley}, {Ferruit}, {Giardino}, {Hainline}, {Hausen}, {Jakobsen}, {Johnson}, {Maiolino}, {Rieke}, {Rieke}, {Rix}, {Robertson}, {Stark}, {Tacchella}, {Williams}, {Willmer}, {Baker}, {Baum}, {Bhatawdekar}, {Boyett}, {Chen}, {Chevallard}, {Circosta}, {Curti}, {Danhaive}, {DeCoursey}, {de Graaff}, {Dressler}, {Egami}, {Helton}, {Hviding}, {Ji}, {Jones}, {Kumari}, {L{\"u}tzgendorf}, {Laseter}, {Looser}, {Lyu}, {Maseda}, {Nelson}, {Parlanti}, {Perna}, {Pusk{\'a}s}, {Rawle}, {Rodr{\'\i}guez Del Pino}, {Sandles}, {Saxena}, {Scholtz}, {Sharpe}, {Shivaei}, {Silcock}, {Simmonds}, {Skarbinski}, {Smit}, {Stone}, {Suess}, {Sun}, {Tang}, {Topping}, {{\"U}bler}, {Villanueva}, {Wallace}, {Whitler}, {Witstok}, \& {Woodrum}}]{eisenstein2023}
{Eisenstein}, D.~J., {Willott}, C., {Alberts}, S., {et~al.} 2023, \href{https://ui.adsabs.harvard.edu/abs/2023arXiv230602465E}{\href{http://dx.doi.org/10.48550/arXiv.2306.02465}{\color{magenta}arXiv e-prints}, arXiv:2306.02465}

\bibitem[{{Fan} {et~al.}(2023){Fan}, {Ba{\~n}ados}, \& {Simcoe}}]{fan2023}
{Fan}, X., {Ba{\~n}ados}, E., \& {Simcoe}, R.~A. 2023, \href{http://dx.doi.org/10.1146/annurev-astro-052920-102455}{\color{magenta}\araa}, \href{https://ui.adsabs.harvard.edu/abs/2023ARA&A..61..373F}{61, 373}

\bibitem[{{Farina} {et~al.}(2022){Farina}, {Schindler}, {Walter}, {Ba{\~n}ados}, {Davies}, {Decarli}, {Eilers}, {Fan}, {Hennawi}, {Mazzucchelli}, {Meyer}, {Trakhtenbrot}, {Volonteri}, {Wang}, {Worseck}, {Yang}, {Gutcke}, {Venemans}, {Bosman}, {Costa}, {De Rosa}, {Drake}, \& {Onoue}}]{farina2022}
{Farina}, E.~P., {Schindler}, J.-T., {Walter}, F., {et~al.} 2022, \href{http://dx.doi.org/10.3847/1538-4357/ac9626}{\color{magenta}\apj}, \href{https://ui.adsabs.harvard.edu/abs/2022ApJ...941..106F}{941, 106}

\bibitem[{{Ferrara} {et~al.}(2023){Ferrara}, {Pallottini}, \& {Dayal}}]{ferrara2023}
{Ferrara}, A., {Pallottini}, A., \& {Dayal}, P. 2023, \href{http://dx.doi.org/10.1093/mnras/stad1095}{\color{magenta}\mnras}, \href{https://ui.adsabs.harvard.edu/abs/2023MNRAS.522.3986F}{522, 3986}

\bibitem[{{Ferrara} {et~al.}(2014){Ferrara}, {Salvadori}, {Yue}, \& {Schleicher}}]{ferrara2014}
{Ferrara}, A., {Salvadori}, S., {Yue}, B., \& {Schleicher}, D. 2014, \href{http://dx.doi.org/10.1093/mnras/stu1280}{\color{magenta}\mnras}, \href{https://ui.adsabs.harvard.edu/abs/2014MNRAS.443.2410F}{443, 2410}

\bibitem[{{Furtak} {et~al.}(2023){Furtak}, {Labb{\'e}}, {Zitrin}, {Greene}, {Dayal}, {Chemerynska}, {Kokorev}, {Miller}, {Goulding}, {Bezanson}, {Brammer}, {Cutler}, {Leja}, {Pan}, {Price}, {Wang}, {Weaver}, {Whitaker}, {Atek}, {Bogd{\'a}n}, {Charlot}, {Curtis-Lake}, {van Dokkum}, {Endsley}, {Fudamoto}, {Fujimoto}, {de Graaff}, {Glazebrook}, {Juneau}, {Marchesini}, {Maseda}, {Nelson}, {Oesch}, {Plat}, {Setton}, {Stark}, \& {Williams}}]{furtak2023}
{Furtak}, L.~J., {Labb{\'e}}, I., {Zitrin}, A., {et~al.} 2023, \href{https://ui.adsabs.harvard.edu/abs/2023arXiv230805735F}{\href{http://dx.doi.org/10.48550/arXiv.2308.05735}{\color{magenta}arXiv e-prints}, arXiv:2308.05735}

\bibitem[{{Gallerani} {et~al.}(2017){Gallerani}, {Fan}, {Maiolino}, \& {Pacucci}}]{gallerani2017}
{Gallerani}, S., {Fan}, X., {Maiolino}, R., \& {Pacucci}, F. 2017, \href{http://dx.doi.org/10.1017/pasa.2017.14}{\color{magenta}\pasa}, \href{https://ui.adsabs.harvard.edu/abs/2017PASA...34...22G}{34, e022}

\bibitem[{{Goulding} {et~al.}(2023){Goulding}, {Greene}, {Setton}, {Labbe}, {Bezanson}, {Miller}, {Atek}, {Bogd{\'a}n}, {Brammer}, {Chemerynska}, {Cutler}, {Dayal}, {Fudamoto}, {Fujimoto}, {Furtak}, {Kokorev}, {Khullar}, {Leja}, {Marchesini}, {Natarajan}, {Nelson}, {Oesch}, {Pan}, {Papovich}, {Price}, {van Dokkum}, {Wang}, {Weaver}, {Whitaker}, \& {Zitrin}}]{goulding2023}
{Goulding}, A.~D., {Greene}, J.~E., {Setton}, D.~J., {et~al.} 2023, \href{http://dx.doi.org/10.3847/2041-8213/acf7c5}{\color{magenta}\apjl}, \href{https://ui.adsabs.harvard.edu/abs/2023ApJ...955L..24G}{955, L24}

\bibitem[{{Greene} {et~al.}(2023){Greene}, {Labbe}, {Goulding}, {Furtak}, {Chemerynska}, {Kokorev}, {Dayal}, {Williams}, {Wang}, {Setton}, {Burgasser}, {Bezanson}, {Atek}, {Brammer}, {Cutler}, {Feldmann}, {Fujimoto}, {Glazebrook}, {de Graaff}, {Leja}, {Marchesini}, {Maseda}, {Matthee}, {Miller}, {Naidu}, {Nanayakkara}, {Oesch}, {Pan}, {Papovich}, {Price}, {van Dokkum}, {Weaver}, {Whitaker}, \& {Zitrin}}]{greene2023}
{Greene}, J.~E., {Labbe}, I., {Goulding}, A.~D., {et~al.} 2023, \href{https://ui.adsabs.harvard.edu/abs/2023arXiv230905714G}{\href{http://dx.doi.org/10.48550/arXiv.2309.05714}{\color{magenta}arXiv e-prints}, arXiv:2309.05714}

\bibitem[{{Greene} {et~al.}(2020){Greene}, {Strader}, \& {Ho}}]{greene2020}
{Greene}, J.~E., {Strader}, J., \& {Ho}, L.~C. 2020, \href{http://dx.doi.org/10.1146/annurev-astro-032620-021835}{\color{magenta}\araa}, \href{https://ui.adsabs.harvard.edu/abs/2020ARA&A..58..257G}{58, 257}

\bibitem[{Gumbel(1958)}]{gumbel1958statistics}
Gumbel, E.~J. 1958, Statistics of extremes (Columbia university press)

\bibitem[{{Habouzit} {et~al.}(2021){Habouzit}, {Li}, {Somerville}, {Genel}, {Pillepich}, {Volonteri}, {Dav{\'e}}, {Rosas-Guevara}, {McAlpine}, {Peirani}, {Hernquist}, {Angl{\'e}s-Alc{\'a}zar}, {Reines}, {Bower}, {Dubois}, {Nelson}, {Pichon}, \& {Vogelsberger}}]{habouzit2021}
{Habouzit}, M., {Li}, Y., {Somerville}, R.~S., {et~al.} 2021, \href{http://dx.doi.org/10.1093/mnras/stab496}{\color{magenta}\mnras}, \href{https://ui.adsabs.harvard.edu/abs/2021MNRAS.503.1940H}{503, 1940}

\bibitem[{{Habouzit} {et~al.}(2022){Habouzit}, {Somerville}, {Li}, {Genel}, {Aird}, {Angl{\'e}s-Alc{\'a}zar}, {Dav{\'e}}, {Georgiev}, {McAlpine}, {Rosas-Guevara}, {Dubois}, {Nelson}, {Banados}, {Hernquist}, {Peirani}, \& {Vogelsberger}}]{habouzit2022}
{Habouzit}, M., {Somerville}, R.~S., {Li}, Y., {et~al.} 2022, \href{http://dx.doi.org/10.1093/mnras/stab3147}{\color{magenta}\mnras}, \href{https://ui.adsabs.harvard.edu/abs/2022MNRAS.509.3015H}{509, 3015}

\bibitem[{{Haehnelt} \& {Rees}(1993)}]{haehnelt1993}
{Haehnelt}, M.~G. \& {Rees}, M.~J. 1993, \href{http://dx.doi.org/10.1093/mnras/263.1.168}{\color{magenta}\mnras}, \href{https://ui.adsabs.harvard.edu/abs/1993MNRAS.263..168H}{263, 168}

\bibitem[{{Hahn} \& {Abel}(2011)}]{hahn2011}
{Hahn}, O. \& {Abel}, T. 2011, \href{http://dx.doi.org/10.1111/j.1365-2966.2011.18820.x}{\color{magenta}\mnras}, \href{https://ui.adsabs.harvard.edu/abs/2011MNRAS.415.2101H}{415, 2101}

\bibitem[{{Harikane} {et~al.}(2023){Harikane}, {Zhang}, {Nakajima}, {Ouchi}, {Isobe}, {Ono}, {Hatano}, {Xu}, \& {Umeda}}]{harikane2023}
{Harikane}, Y., {Zhang}, Y., {Nakajima}, K., {et~al.} 2023, \href{http://dx.doi.org/10.3847/1538-4357/ad029e}{\color{magenta}\apj}, \href{https://ui.adsabs.harvard.edu/abs/2023ApJ...959...39H}{959, 39}

\bibitem[{{Harrison} \& {Coles}(2011)}]{harrison2011}
{Harrison}, I. \& {Coles}, P. 2011, \href{http://dx.doi.org/10.1111/j.1745-3933.2011.01134.x}{\color{magenta}\mnras}, \href{https://ui.adsabs.harvard.edu/abs/2011MNRAS.418L..20H}{418, L20}

\bibitem[{{Hoyle} \& {Lyttleton}(1939)}]{hoyle1939}
{Hoyle}, F. \& {Lyttleton}, R.~A. 1939, \href{http://dx.doi.org/10.1017/S0305004100021150}{\color{magenta}Proceedings of the Cambridge Philosophical Society}, \href{https://ui.adsabs.harvard.edu/abs/1939PCPS...35..405H}{35, 405}

\bibitem[{{Inayoshi} {et~al.}(2020){Inayoshi}, {Visbal}, \& {Haiman}}]{inayoshi2020}
{Inayoshi}, K., {Visbal}, E., \& {Haiman}, Z. 2020, \href{http://dx.doi.org/10.1146/annurev-astro-120419-014455}{\color{magenta}\araa}, \href{https://ui.adsabs.harvard.edu/abs/2020ARA&A..58...27I}{58, 27}

\bibitem[{{Jeon} {et~al.}(2024){Jeon}, {Bromm}, {Liu}, \& {Finkelstein}}]{jeon2024}
{Jeon}, J., {Bromm}, V., {Liu}, B., \& {Finkelstein}, S.~L. 2024, \href{https://ui.adsabs.harvard.edu/abs/2024arXiv240218773J}{\href{http://dx.doi.org/10.48550/arXiv.2402.18773}{\color{magenta}arXiv e-prints}, arXiv:2402.18773}

\bibitem[{{Kokorev} {et~al.}(2023){Kokorev}, {Fujimoto}, {Labbe}, {Greene}, {Bezanson}, {Dayal}, {Nelson}, {Atek}, {Brammer}, {Caputi}, {Chemerynska}, {Cutler}, {Feldmann}, {Fudamoto}, {Furtak}, {Goulding}, {de Graaff}, {Leja}, {Marchesini}, {Miller}, {Nanayakkara}, {Oesch}, {Pan}, {Price}, {Setton}, {Smit}, {Stefanon}, {Wang}, {Weaver}, {Whitaker}, {Williams}, \& {Zitrin}}]{kokorev2023}
{Kokorev}, V., {Fujimoto}, S., {Labbe}, I., {et~al.} 2023, \href{http://dx.doi.org/10.3847/2041-8213/ad037a}{\color{magenta}\apjl}, \href{https://ui.adsabs.harvard.edu/abs/2023ApJ...957L...7K}{957, L7}

\bibitem[{Kolmogorov(1933)}]{kolmogorov1933sulla}
Kolmogorov, A. 1933, Inst. Ital. Attuari, Giorn., 4, 83

\bibitem[{Kotz \& Nadarajah(2000)}]{kotz2000extreme}
Kotz, S. \& Nadarajah, S. 2000, Extreme value distributions: theory and applications (world scientific)

\bibitem[{{Kroupa} {et~al.}(2020){Kroupa}, {Subr}, {Jerabkova}, \& {Wang}}]{kroupa2020}
{Kroupa}, P., {Subr}, L., {Jerabkova}, T., \& {Wang}, L. 2020, \href{http://dx.doi.org/10.1093/mnras/staa2276}{\color{magenta}\mnras}, \href{https://ui.adsabs.harvard.edu/abs/2020MNRAS.498.5652K}{498, 5652}

\bibitem[{{Krumholz}(2017)}]{krumholz2017book}
{Krumholz}, M.~R. 2017, {Star Formation}

\bibitem[{{Larson} {et~al.}(2023){Larson}, {Finkelstein}, {Kocevski}, {Hutchison}, {Trump}, {Arrabal Haro}, {Bromm}, {Cleri}, {Dickinson}, {Fujimoto}, {Kartaltepe}, {Koekemoer}, {Papovich}, {Pirzkal}, {Tacchella}, {Zavala}, {Bagley}, {Behroozi}, {Champagne}, {Cole}, {Jung}, {Morales}, {Yang}, {Zhang}, {Zitrin}, {Amor{\'\i}n}, {Burgarella}, {Casey}, {Ch{\'a}vez Ortiz}, {Cox}, {Chworowsky}, {Fontana}, {Gawiser}, {Grazian}, {Grogin}, {Harish}, {Hathi}, {Hirschmann}, {Holwerda}, {Juneau}, {Leung}, {Lucas}, {McGrath}, {P{\'e}rez-Gonz{\'a}lez}, {Rigby}, {Seill{\'e}}, {Simons}, {de La Vega}, {Weiner}, {Wilkins}, {Yung}, \& {Ceers Team}}]{larson2023}
{Larson}, R.~L., {Finkelstein}, S.~L., {Kocevski}, D.~D., {et~al.} 2023, \href{http://dx.doi.org/10.3847/2041-8213/ace619}{\color{magenta}\apjl}, \href{https://ui.adsabs.harvard.edu/abs/2023ApJ...953L..29L}{953, L29}

\bibitem[{{Latif} \& {Ferrara}(2016)}]{latif2016}
{Latif}, M.~A. \& {Ferrara}, A. 2016, \href{http://dx.doi.org/10.1017/pasa.2016.41}{\color{magenta}\pasa}, \href{https://ui.adsabs.harvard.edu/abs/2016PASA...33...51L}{33, e051}

\bibitem[{{Le F{\`e}vre} {et~al.}(2019){Le F{\`e}vre}, {Lemaux}, {Nakajima}, {Schaerer}, {Talia}, {Zamorani}, {Cassata}, {Garilli}, {Maccagni}, {Pentericci}, {Tasca}, {Zucca}, {Amorin}, {Bardelli}, {Cimatti}, {Giavalisco}, {Guaita}, {Hathi}, {Marchi}, {Vanzella}, {Vergani}, \& {Dunlop}}]{lefevre2019}
{Le F{\`e}vre}, O., {Lemaux}, B.~C., {Nakajima}, K., {et~al.} 2019, \href{http://dx.doi.org/10.1051/0004-6361/201732197}{\color{magenta}\aap}, \href{https://ui.adsabs.harvard.edu/abs/2019A&A...625A..51L}{625, A51}

\bibitem[{{Lovell} {et~al.}(2023){Lovell}, {Harrison}, {Harikane}, {Tacchella}, \& {Wilkins}}]{lovell2023}
{Lovell}, C.~C., {Harrison}, I., {Harikane}, Y., {Tacchella}, S., \& {Wilkins}, S.~M. 2023, \href{http://dx.doi.org/10.1093/mnras/stac3224}{\color{magenta}\mnras}, \href{https://ui.adsabs.harvard.edu/abs/2023MNRAS.518.2511L}{518, 2511}

\bibitem[{{Lupi} {et~al.}(2016){Lupi}, {Haardt}, {Dotti}, {Fiacconi}, {Mayer}, \& {Madau}}]{lupi2016}
{Lupi}, A., {Haardt}, F., {Dotti}, M., {et~al.} 2016, \href{http://dx.doi.org/10.1093/mnras/stv2877}{\color{magenta}\mnras}, \href{https://ui.adsabs.harvard.edu/abs/2016MNRAS.456.2993L}{456, 2993}

\bibitem[{{Lupi} {et~al.}(2023){Lupi}, {Quadri}, {Volonteri}, {Colpi}, \& {Regan}}]{lupi2023}
{Lupi}, A., {Quadri}, G., {Volonteri}, M., {Colpi}, M., \& {Regan}, J.~A. 2023, \href{https://ui.adsabs.harvard.edu/abs/2023arXiv231208422L}{\href{http://dx.doi.org/10.48550/arXiv.2312.08422}{\color{magenta}arXiv e-prints}, arXiv:2312.08422}

\bibitem[{{Madau} {et~al.}(2014){Madau}, {Haardt}, \& {Dotti}}]{madau2014}
{Madau}, P., {Haardt}, F., \& {Dotti}, M. 2014, \href{http://dx.doi.org/10.1088/2041-8205/784/2/L38}{\color{magenta}\apjl}, \href{https://ui.adsabs.harvard.edu/abs/2014ApJ...784L..38M}{784, L38}

\bibitem[{{Madau} \& {Rees}(2001)}]{madau2001}
{Madau}, P. \& {Rees}, M.~J. 2001, \href{http://dx.doi.org/10.1086/319848}{\color{magenta}\apjl}, \href{https://ui.adsabs.harvard.edu/abs/2001ApJ...551L..27M}{551, L27}

\bibitem[{{Maiolino} {et~al.}(2023{\natexlab{a}}){Maiolino}, {Scholtz}, {Curtis-Lake}, {Carniani}, {Baker}, {de Graaff}, {Tacchella}, {{\"U}bler}, {D'Eugenio}, {Witstok}, {Curti}, {Arribas}, {Bunker}, {Charlot}, {Chevallard}, {Eisenstein}, {Egami}, {Ji}, {Jones}, {Lyu}, {Rawle}, {Robertson}, {Rujopakarn}, {Perna}, {Sun}, {Venturi}, {Williams}, \& {Willott}}]{maiolino2023b}
{Maiolino}, R., {Scholtz}, J., {Curtis-Lake}, E., {et~al.} 2023{\natexlab{a}}, \href{https://ui.adsabs.harvard.edu/abs/2023arXiv230801230M}{\href{http://dx.doi.org/10.48550/arXiv.2308.01230}{\color{magenta}arXiv e-prints}, arXiv:2308.01230}

\bibitem[{{Maiolino} {et~al.}(2023{\natexlab{b}}){Maiolino}, {Scholtz}, {Witstok}, {Carniani}, {D'Eugenio}, {de Graaff}, {Uebler}, {Tacchella}, {Curtis-Lake}, {Arribas}, {Bunker}, {Charlot}, {Chevallard}, {Curti}, {Looser}, {Maseda}, {Rawle}, {Rodriguez Del Pino}, {Willott}, {Egami}, {Eisenstein}, {Hainline}, {Robertson}, {Williams}, {Willmer}, {Baker}, {Boyett}, {DeCoursey}, {Fabian}, {Helton}, {Ji}, {Jones}, {Kumari}, {Laporte}, {Nelson}, {Perna}, {Sandles}, {Shivaei}, \& {Sun}}]{maiolino2023a}
{Maiolino}, R., {Scholtz}, J., {Witstok}, J., {et~al.} 2023{\natexlab{b}}, \href{https://ui.adsabs.harvard.edu/abs/2023arXiv230512492M}{\href{http://dx.doi.org/10.48550/arXiv.2305.12492}{\color{magenta}arXiv e-prints}, arXiv:2305.12492}

\bibitem[{{Massonneau} {et~al.}(2023){Massonneau}, {Volonteri}, {Dubois}, \& {Beckmann}}]{massonneau2023}
{Massonneau}, W., {Volonteri}, M., {Dubois}, Y., \& {Beckmann}, R.~S. 2023, \href{http://dx.doi.org/10.1051/0004-6361/202243170}{\color{magenta}\aap}, \href{https://ui.adsabs.harvard.edu/abs/2023A&A...670A.180M}{670, A180}

\bibitem[{{Mayer} \& {Bonoli}(2019)}]{mayer2019}
{Mayer}, L. \& {Bonoli}, S. 2019, \href{http://dx.doi.org/10.1088/1361-6633/aad6a5}{\color{magenta}Reports on Progress in Physics}, \href{https://ui.adsabs.harvard.edu/abs/2019RPPh...82a6901M}{82, 016901}

\bibitem[{{Mazzucchelli} {et~al.}(2023){Mazzucchelli}, {Bischetti}, {D'Odorico}, {Feruglio}, {Schindler}, {Onoue}, {Ba{\~n}ados}, {Becker}, {Bian}, {Carniani}, {Decarli}, {Eilers}, {Farina}, {Gallerani}, {Lai}, {Meyer}, {Rojas-Ruiz}, {Satyavolu}, {Venemans}, {Wang}, {Yang}, \& {Zhu}}]{mazzucchelli2023}
{Mazzucchelli}, C., {Bischetti}, M., {D'Odorico}, V., {et~al.} 2023, \href{http://dx.doi.org/10.1051/0004-6361/202346317}{\color{magenta}\aap}, \href{https://ui.adsabs.harvard.edu/abs/2023A&A...676A..71M}{676, A71}

\bibitem[{{Mikelsons} {et~al.}(2009){Mikelsons}, {Silk}, \& {Zuntz}}]{mikelsons2009}
{Mikelsons}, G., {Silk}, J., \& {Zuntz}, J. 2009, \href{http://dx.doi.org/10.1111/j.1365-2966.2009.15503.x}{\color{magenta}\mnras}, \href{https://ui.adsabs.harvard.edu/abs/2009MNRAS.400..898M}{400, 898}

\bibitem[{{Naidu} {et~al.}(2022){Naidu}, {Oesch}, {van Dokkum}, {Nelson}, {Suess}, {Brammer}, {Whitaker}, {Illingworth}, {Bouwens}, {Tacchella}, {Matthee}, {Allen}, {Bezanson}, {Conroy}, {Labbe}, {Leja}, {Leonova}, {Magee}, {Price}, {Setton}, {Strait}, {Stefanon}, {Toft}, {Weaver}, \& {Weibel}}]{naidu2022}
{Naidu}, R.~P., {Oesch}, P.~A., {van Dokkum}, P., {et~al.} 2022, \href{http://dx.doi.org/10.3847/2041-8213/ac9b22}{\color{magenta}\apjl}, \href{https://ui.adsabs.harvard.edu/abs/2022ApJ...940L..14N}{940, L14}

\bibitem[{{Oesch} {et~al.}(2016){Oesch}, {Brammer}, {van Dokkum}, {Illingworth}, {Bouwens}, {Labb{\'e}}, {Franx}, {Momcheva}, {Ashby}, {Fazio}, {Gonzalez}, {Holden}, {Magee}, {Skelton}, {Smit}, {Spitler}, {Trenti}, \& {Willner}}]{oesch2016}
{Oesch}, P.~A., {Brammer}, G., {van Dokkum}, P.~G., {et~al.} 2016, \href{http://dx.doi.org/10.3847/0004-637X/819/2/129}{\color{magenta}\apj}, \href{https://ui.adsabs.harvard.edu/abs/2016ApJ...819..129O}{819, 129}

\bibitem[{{Orofino} {et~al.}(2018){Orofino}, {Ferrara}, \& {Gallerani}}]{orofino2018}
{Orofino}, M.~C., {Ferrara}, A., \& {Gallerani}, S. 2018, \href{http://dx.doi.org/10.1093/mnras/sty1482}{\color{magenta}\mnras}, \href{https://ui.adsabs.harvard.edu/abs/2018MNRAS.480..681O}{480, 681}

\bibitem[{{Pacucci} \& {Ferrara}(2015)}]{pacucci2015}
{Pacucci}, F. \& {Ferrara}, A. 2015, \href{http://dx.doi.org/10.1093/mnras/stv018}{\color{magenta}\mnras}, \href{https://ui.adsabs.harvard.edu/abs/2015MNRAS.448..104P}{448, 104}

\bibitem[{{Pacucci} {et~al.}(2017){Pacucci}, {Natarajan}, {Volonteri}, {Cappelluti}, \& {Urry}}]{pacucci2017}
{Pacucci}, F., {Natarajan}, P., {Volonteri}, M., {Cappelluti}, N., \& {Urry}, C.~M. 2017, \href{http://dx.doi.org/10.3847/2041-8213/aa9aea}{\color{magenta}\apjl}, \href{https://ui.adsabs.harvard.edu/abs/2017ApJ...850L..42P}{850, L42}

\bibitem[{{Pacucci} {et~al.}(2023){Pacucci}, {Nguyen}, {Carniani}, {Maiolino}, \& {Fan}}]{pacucci2023}
{Pacucci}, F., {Nguyen}, B., {Carniani}, S., {Maiolino}, R., \& {Fan}, X. 2023, \href{http://dx.doi.org/10.3847/2041-8213/ad0158}{\color{magenta}\apjl}, \href{https://ui.adsabs.harvard.edu/abs/2023ApJ...957L...3P}{957, L3}

\bibitem[{{Pensabene} {et~al.}(2020){Pensabene}, {Carniani}, {Perna}, {Cresci}, {Decarli}, {Maiolino}, \& {Marconi}}]{pensabene2020}
{Pensabene}, A., {Carniani}, S., {Perna}, M., {et~al.} 2020, \href{http://dx.doi.org/10.1051/0004-6361/201936634}{\color{magenta}\aap}, \href{https://ui.adsabs.harvard.edu/abs/2020A&A...637A..84P}{637, A84}

\bibitem[{{Pezzulli} {et~al.}(2016){Pezzulli}, {Valiante}, \& {Schneider}}]{pezzulli2016}
{Pezzulli}, E., {Valiante}, R., \& {Schneider}, R. 2016, \href{http://dx.doi.org/10.1093/mnras/stw505}{\color{magenta}\mnras}, \href{https://ui.adsabs.harvard.edu/abs/2016MNRAS.458.3047P}{458, 3047}

\bibitem[{{Pezzulli} {et~al.}(2017){Pezzulli}, {Volonteri}, {Schneider}, \& {Valiante}}]{pezzulli2017}
{Pezzulli}, E., {Volonteri}, M., {Schneider}, R., \& {Valiante}, R. 2017, \href{http://dx.doi.org/10.1093/mnras/stx1640}{\color{magenta}\mnras}, \href{https://ui.adsabs.harvard.edu/abs/2017MNRAS.471..589P}{471, 589}

\bibitem[{{Planck Collaboration} {et~al.}(2020){Planck Collaboration}, {Aghanim}, {Akrami}, {Ashdown}, {Aumont}, {Baccigalupi}, {Ballardini}, {Banday}, {Barreiro}, {Bartolo}, {Basak}, {Battye}, {Benabed}, {Bernard}, {Bersanelli}, {Bielewicz}, {Bock}, {Bond}, {Borrill}, {Bouchet}, {Boulanger}, {Bucher}, {Burigana}, {Butler}, {Calabrese}, {Cardoso}, {Carron}, {Challinor}, {Chiang}, {Chluba}, {Colombo}, {Combet}, {Contreras}, {Crill}, {Cuttaia}, {de Bernardis}, {de Zotti}, {Delabrouille}, {Delouis}, {Di Valentino}, {Diego}, {Dor{\'e}}, {Douspis}, {Ducout}, {Dupac}, {Dusini}, {Efstathiou}, {Elsner}, {En{\ss}lin}, {Eriksen}, {Fantaye}, {Farhang}, {Fergusson}, {Fernandez-Cobos}, {Finelli}, {Forastieri}, {Frailis}, {Fraisse}, {Franceschi}, {Frolov}, {Galeotta}, {Galli}, {Ganga}, {G{\'e}nova-Santos}, {Gerbino}, {Ghosh}, {Gonz{\'a}lez-Nuevo}, {G{\'o}rski}, {Gratton}, {Gruppuso}, {Gudmundsson}, {Hamann}, {Handley}, {Hansen}, {Herranz}, {Hildebrandt}, {Hivon}, {Huang}, {Jaffe}, {Jones}, {Karakci}, {Keih{\"a}nen},
  {Keskitalo}, {Kiiveri}, {Kim}, {Kisner}, {Knox}, {Krachmalnicoff}, {Kunz}, {Kurki-Suonio}, {Lagache}, {Lamarre}, {Lasenby}, {Lattanzi}, {Lawrence}, {Le Jeune}, {Lemos}, {Lesgourgues}, {Levrier}, {Lewis}, {Liguori}, {Lilje}, {Lilley}, {Lindholm}, {L{\'o}pez-Caniego}, {Lubin}, {Ma}, {Mac{\'\i}as-P{\'e}rez}, {Maggio}, {Maino}, {Mandolesi}, {Mangilli}, {Marcos-Caballero}, {Maris}, {Martin}, {Martinelli}, {Mart{\'\i}nez-Gonz{\'a}lez}, {Matarrese}, {Mauri}, {McEwen}, {Meinhold}, {Melchiorri}, {Mennella}, {Migliaccio}, {Millea}, {Mitra}, {Miville-Desch{\^e}nes}, {Molinari}, {Montier}, {Morgante}, {Moss}, {Natoli}, {N{\o}rgaard-Nielsen}, {Pagano}, {Paoletti}, {Partridge}, {Patanchon}, {Peiris}, {Perrotta}, {Pettorino}, {Piacentini}, {Polastri}, {Polenta}, {Puget}, {Rachen}, {Reinecke}, {Remazeilles}, {Renzi}, {Rocha}, {Rosset}, {Roudier}, {Rubi{\~n}o-Mart{\'\i}n}, {Ruiz-Granados}, {Salvati}, {Sandri}, {Savelainen}, {Scott}, {Shellard}, {Sirignano}, {Sirri}, {Spencer}, {Sunyaev}, {Suur-Uski}, {Tauber}, {Tavagnacco},
  {Tenti}, {Toffolatti}, {Tomasi}, {Trombetti}, {Valenziano}, {Valiviita}, {Van Tent}, {Vibert}, {Vielva}, {Villa}, {Vittorio}, {Wandelt}, {Wehus}, {White}, {White}, {Zacchei}, \& {Zonca}}]{planck2020}
{Planck Collaboration}, {Aghanim}, N., {Akrami}, Y., {et~al.} 2020, \href{http://dx.doi.org/10.1051/0004-6361/201833910}{\color{magenta}\aap}, \href{https://ui.adsabs.harvard.edu/abs/2020A&A...641A...6P}{641, A6}

\bibitem[{{Reines} \& {Volonteri}(2015)}]{reines2015}
{Reines}, A.~E. \& {Volonteri}, M. 2015, \href{http://dx.doi.org/10.1088/0004-637X/813/2/82}{\color{magenta}\apj}, \href{https://ui.adsabs.harvard.edu/abs/2015ApJ...813...82R}{813, 82}

\bibitem[{{Richards} {et~al.}(2006){Richards}, {Haiman}, {Pindor}, {Strauss}, {Fan}, {Eisenstein}, {Schneider}, {Bahcall}, {Brinkmann}, \& {Fukugita}}]{richards2006}
{Richards}, G.~T., {Haiman}, Z., {Pindor}, B., {et~al.} 2006, \href{http://dx.doi.org/10.1086/498063}{\color{magenta}\aj}, \href{https://ui.adsabs.harvard.edu/abs/2006AJ....131...49R}{131, 49}

\bibitem[{{Rodr{\'\i}guez Hidalgo}(2009)}]{rodriguez2009phd}
{Rodr{\'\i}guez Hidalgo}, P. 2009, \href{https://ui.adsabs.harvard.edu/abs/2009PhDT.......203R}{{High velocity outflows in quasars}}, PhD thesis, University of Florida

\bibitem[{{Schneider} {et~al.}(2023){Schneider}, {Valiante}, {Trinca}, {Graziani}, {Volonteri}, \& {Maiolino}}]{schneider2023}
{Schneider}, R., {Valiante}, R., {Trinca}, A., {et~al.} 2023, \href{http://dx.doi.org/10.1093/mnras/stad2503}{\color{magenta}\mnras}, \href{https://ui.adsabs.harvard.edu/abs/2023MNRAS.526.3250S}{526, 3250}

\bibitem[{{Scholtz} {et~al.}(2023){Scholtz}, {Witten}, {Laporte}, {Ubler}, {Perna}, {Maiolino}, {Arribas}, {Baker}, {Bennett}, {D'Eugenio}, {Tacchella}, {Witstok}, {Bunker}, {Carniani}, {Charlot}, {Curtis-Lake}, {Eisenstein}, {Robertson}, {Rodriguez Del Pino}, {Simmonds}, {Smit}, {Venturi}, {Williams}, \& {Willmer}}]{scholtz2023}
{Scholtz}, J., {Witten}, C., {Laporte}, N., {et~al.} 2023, \href{https://ui.adsabs.harvard.edu/abs/2023arXiv230609142S}{\href{http://dx.doi.org/10.48550/arXiv.2306.09142}{\color{magenta}arXiv e-prints}, arXiv:2306.09142}

\bibitem[{{Shen} \& {Liu}(2012)}]{Shen_Liu2012}
{Shen}, Y. \& {Liu}, X. 2012, \href{http://dx.doi.org/10.1088/0004-637X/753/2/125}{\color{magenta}\apj}, \href{https://ui.adsabs.harvard.edu/abs/2012ApJ...753..125S}{753, 125}

\bibitem[{{Shen} {et~al.}(2011){Shen}, {Richards}, {Strauss}, {Hall}, {Schneider}, {Snedden}, {Bizyaev}, {Brewington}, {Malanushenko}, {Malanushenko}, {Oravetz}, {Pan}, \& {Simmons}}]{shen2011}
{Shen}, Y., {Richards}, G.~T., {Strauss}, M.~A., {et~al.} 2011, \href{http://dx.doi.org/10.1088/0067-0049/194/2/45}{\color{magenta}\apjs}, \href{https://ui.adsabs.harvard.edu/abs/2011ApJS..194...45S}{194, 45}

\bibitem[{{Sheth} \& {Tormen}(1999)}]{sheth1999}
{Sheth}, R.~K. \& {Tormen}, G. 1999, \href{http://dx.doi.org/10.1046/j.1365-8711.1999.02692.x}{\color{magenta}\mnras}, \href{https://ui.adsabs.harvard.edu/abs/1999MNRAS.308..119S}{308, 119}

\bibitem[{{S{\k{a}}dowski}(2009)}]{sadowski2009}
{S{\k{a}}dowski}, A. 2009, \href{http://dx.doi.org/10.1088/0067-0049/183/2/171}{\color{magenta}\apjs}, \href{https://ui.adsabs.harvard.edu/abs/2009ApJS..183..171S}{183, 171}

\bibitem[{{S{\k{a}}dowski} \& {Gaspari}(2017)}]{sadowski2017}
{S{\k{a}}dowski}, A. \& {Gaspari}, M. 2017, \href{http://dx.doi.org/10.1093/mnras/stx543}{\color{magenta}\mnras}, \href{https://ui.adsabs.harvard.edu/abs/2017MNRAS.468.1398S}{468, 1398}

\bibitem[{Smirnov(1948)}]{smirnov1948table}
Smirnov, N. 1948, The annals of mathematical statistics, 19, 279

\bibitem[{{Smith} {et~al.}(2018){Smith}, {Regan}, {Downes}, {Norman}, {O'Shea}, \& {Wise}}]{smith2018}
{Smith}, B.~D., {Regan}, J.~A., {Downes}, T.~P., {et~al.} 2018, \href{http://dx.doi.org/10.1093/mnras/sty2103}{\color{magenta}\mnras}, \href{https://ui.adsabs.harvard.edu/abs/2018MNRAS.480.3762S}{480, 3762}

\bibitem[{{Springel}(2005)}]{springel2005}
{Springel}, V. 2005, \href{http://dx.doi.org/10.1111/j.1365-2966.2005.09655.x}{\color{magenta}\mnras}, \href{https://ui.adsabs.harvard.edu/abs/2005MNRAS.364.1105S}{364, 1105}

\bibitem[{{Tacchella} {et~al.}(2023){Tacchella}, {Eisenstein}, {Hainline}, {Johnson}, {Baker}, {Helton}, {Robertson}, {Suess}, {Chen}, {Nelson}, {Pusk{\'a}s}, {Sun}, {Alberts}, {Egami}, {Hausen}, {Rieke}, {Rieke}, {Shivaei}, {Williams}, {Willmer}, {Bunker}, {Cameron}, {Carniani}, {Charlot}, {Curti}, {Curtis-Lake}, {Looser}, {Maiolino}, {Maseda}, {Rawle}, {Rix}, {Smit}, {{\"U}bler}, {Willott}, {Witstok}, {Baum}, {Bhatawdekar}, {Boyett}, {Danhaive}, {de Graaff}, {Endsley}, {Ji}, {Lyu}, {Sandles}, {Saxena}, {Scholtz}, {Topping}, \& {Whitler}}]{tacchella2023}
{Tacchella}, S., {Eisenstein}, D.~J., {Hainline}, K., {et~al.} 2023, \href{http://dx.doi.org/10.3847/1538-4357/acdbc6}{\color{magenta}\apj}, \href{https://ui.adsabs.harvard.edu/abs/2023ApJ...952...74T}{952, 74}

\bibitem[{{Terao} {et~al.}(2022){Terao}, {Nagao}, {Onishi}, {Matsuoka}, {Akiyama}, {Matsuoka}, \& {Yamashita}}]{terao2022}
{Terao}, K., {Nagao}, T., {Onishi}, K., {et~al.} 2022, \href{http://dx.doi.org/10.3847/1538-4357/ac5b71}{\color{magenta}\apj}, \href{https://ui.adsabs.harvard.edu/abs/2022ApJ...929...51T}{929, 51}

\bibitem[{{Trakhtenbrot} \& {Netzer}(2012)}]{trakhtenbrot2012}
{Trakhtenbrot}, B. \& {Netzer}, H. 2012, \href{http://dx.doi.org/10.1111/j.1365-2966.2012.22056.x}{\color{magenta}\mnras}, \href{https://ui.adsabs.harvard.edu/abs/2012MNRAS.427.3081T}{427, 3081}

\bibitem[{{Trinca} {et~al.}(2023){Trinca}, {Schneider}, {Valiante}, {Graziani}, {Ferrotti}, {Omukai}, \& {Chon}}]{trinca2023}
{Trinca}, A., {Schneider}, R., {Valiante}, R., {et~al.} 2023, \href{https://ui.adsabs.harvard.edu/abs/2023arXiv230504944T}{\href{http://dx.doi.org/10.48550/arXiv.2305.04944}{\color{magenta}arXiv e-prints}, arXiv:2305.04944}

\bibitem[{{Trinca} {et~al.}(2022){Trinca}, {Schneider}, {Valiante}, {Graziani}, {Zappacosta}, \& {Shankar}}]{trinca2022}
{Trinca}, A., {Schneider}, R., {Valiante}, R., {et~al.} 2022, \href{http://dx.doi.org/10.1093/mnras/stac062}{\color{magenta}\mnras}, \href{https://ui.adsabs.harvard.edu/abs/2022MNRAS.511..616T}{511, 616}

\bibitem[{{Valentini} {et~al.}(2019){Valentini}, {Borgani}, {Bressan}, {Murante}, {Tornatore}, \& {Monaco}}]{valentini2019}
{Valentini}, M., {Borgani}, S., {Bressan}, A., {et~al.} 2019, \href{http://dx.doi.org/10.1093/mnras/stz492}{\color{magenta}\mnras}, \href{https://ui.adsabs.harvard.edu/abs/2019MNRAS.485.1384V}{485, 1384}

\bibitem[{{Valentini} {et~al.}(2021){Valentini}, {Gallerani}, \& {Ferrara}}]{Valentini:2021}
{Valentini}, M., {Gallerani}, S., \& {Ferrara}, A. 2021, \href{http://dx.doi.org/10.1093/mnras/stab1992}{\color{magenta}\mnras}, \href{https://ui.adsabs.harvard.edu/abs/2021MNRAS.507....1V}{507, 1}

\bibitem[{{Valentini} {et~al.}(2020){Valentini}, {Murante}, {Borgani}, {Granato}, {Monaco}, {Brighenti}, {Tornatore}, {Bressan}, \& {Lapi}}]{valentini2020}
{Valentini}, M., {Murante}, G., {Borgani}, S., {et~al.} 2020, \href{http://dx.doi.org/10.1093/mnras/stz3131}{\color{magenta}\mnras}, \href{https://ui.adsabs.harvard.edu/abs/2020MNRAS.491.2779V}{491, 2779}

\bibitem[{{Valentini} {et~al.}(2017){Valentini}, {Murante}, {Borgani}, {Monaco}, {Bressan}, \& {Beck}}]{valentini2017}
{Valentini}, M., {Murante}, G., {Borgani}, S., {et~al.} 2017, \href{http://dx.doi.org/10.1093/mnras/stx1352}{\color{magenta}\mnras}, \href{https://ui.adsabs.harvard.edu/abs/2017MNRAS.470.3167V}{470, 3167}

\bibitem[{{Venemans} {et~al.}(2017){Venemans}, {Walter}, {Decarli}, {Ba{\~n}ados}, {Carilli}, {Winters}, {Schuster}, {da Cunha}, {Fan}, {Farina}, {Mazzucchelli}, {Rix}, \& {Weiss}}]{venemans2017}
{Venemans}, B.~P., {Walter}, F., {Decarli}, R., {et~al.} 2017, \href{http://dx.doi.org/10.3847/2041-8213/aa943a}{\color{magenta}\apjl}, \href{https://ui.adsabs.harvard.edu/abs/2017ApJ...851L...8V}{851, L8}

\bibitem[{{Venemans} {et~al.}(2020){Venemans}, {Walter}, {Neeleman}, {Novak}, {Otter}, {Decarli}, {Ba{\~n}ados}, {Drake}, {Farina}, {Kaasinen}, {Mazzucchelli}, {Carilli}, {Fan}, {Rix}, \& {Wang}}]{venemans2020}
{Venemans}, B.~P., {Walter}, F., {Neeleman}, M., {et~al.} 2020, \href{http://dx.doi.org/10.3847/1538-4357/abc563}{\color{magenta}\apj}, \href{https://ui.adsabs.harvard.edu/abs/2020ApJ...904..130V}{904, 130}

\bibitem[{{Vestergaard} \& {Osmer}(2009)}]{vestergaard2009}
{Vestergaard}, M. \& {Osmer}, P.~S. 2009, \href{http://dx.doi.org/10.1088/0004-637X/699/1/800}{\color{magenta}\apj}, \href{https://ui.adsabs.harvard.edu/abs/2009ApJ...699..800V}{699, 800}

\bibitem[{{Vestergaard} \& {Peterson}(2006)}]{vestergaard2006}
{Vestergaard}, M. \& {Peterson}, B.~M. 2006, \href{http://dx.doi.org/10.1086/500572}{\color{magenta}\apj}, \href{https://ui.adsabs.harvard.edu/abs/2006ApJ...641..689V}{641, 689}

\bibitem[{{Volonteri} \& {Bellovary}(2012)}]{volonteri2012}
{Volonteri}, M. \& {Bellovary}, J. 2012, \href{http://dx.doi.org/10.1088/0034-4885/75/12/124901}{\color{magenta}Reports on Progress in Physics}, \href{https://ui.adsabs.harvard.edu/abs/2012RPPh...75l4901V}{75, 124901}

\bibitem[{{Volonteri} {et~al.}(2021){Volonteri}, {Habouzit}, \& {Colpi}}]{volonteri2021}
{Volonteri}, M., {Habouzit}, M., \& {Colpi}, M. 2021, \href{http://dx.doi.org/10.1038/s42254-021-00364-9}{\color{magenta}Nature Reviews Physics}, \href{https://ui.adsabs.harvard.edu/abs/2021NatRP...3..732V}{3, 732}

\bibitem[{{Waizmann} {et~al.}(2011){Waizmann}, {Ettori}, \& {Moscardini}}]{waizmann2011}
{Waizmann}, J.~C., {Ettori}, S., \& {Moscardini}, L. 2011, \href{http://dx.doi.org/10.1111/j.1365-2966.2011.19496.x}{\color{magenta}\mnras}, \href{https://ui.adsabs.harvard.edu/abs/2011MNRAS.418..456W}{418, 456}

\bibitem[{{Wang} {et~al.}(2021){Wang}, {Fan}, {Yang}, {Mazzucchelli}, {Wu}, {Li}, {Ba{\~n}ados}, {Farina}, {Nanni}, {Ai}, {Bian}, {Davies}, {Decarli}, {Hennawi}, {Schindler}, {Venemans}, \& {Walter}}]{wang2021}
{Wang}, F., {Fan}, X., {Yang}, J., {et~al.} 2021, \href{http://dx.doi.org/10.3847/1538-4357/abcc5e}{\color{magenta}\apj}, \href{https://ui.adsabs.harvard.edu/abs/2021ApJ...908...53W}{908, 53}

\bibitem[{{Wang} {et~al.}(2016){Wang}, {Wu}, {Neri}, {Fan}, {Walter}, {Carilli}, {Momjian}, {Bertoldi}, {Strauss}, {Li}, {Wang}, {Riechers}, {Jiang}, {Omont}, {Wagg}, \& {Cox}}]{wang2016}
{Wang}, R., {Wu}, X.-B., {Neri}, R., {et~al.} 2016, \href{http://dx.doi.org/10.3847/0004-637X/830/1/53}{\color{magenta}\apj}, \href{https://ui.adsabs.harvard.edu/abs/2016ApJ...830...53W}{830, 53}

\bibitem[{{Yang} {et~al.}(2023){Yang}, {Wang}, {Fan}, {Hennawi}, {Barth}, {Ba{\~n}ados}, {Sun}, {Liu}, {Cai}, {Jiang}, {Li}, {Onoue}, {Schindler}, {Shen}, {Wu}, {Bhowmick}, {Bieri}, {Blecha}, {Bosman}, {Champagne}, {Colina}, {Connor}, {Costa}, {Davies}, {Decarli}, {De Rosa}, {Drake}, {Egami}, {Eilers}, {Evans}, {Farina}, {Habouzit}, {Haiman}, {Jin}, {Jun}, {Kakiichi}, {Khusanova}, {Kulkarni}, {Loiacono}, {Lupi}, {Mazzucchelli}, {Pan}, {Rojas-Ruiz}, {Strauss}, {Tee}, {Trakhtenbrot}, {Trebitsch}, {Venemans}, {Vestergaard}, {Volonteri}, {Walter}, {Xie}, {Yue}, {Zhang}, {Zhang}, \& {Zou}}]{yang2023}
{Yang}, J., {Wang}, F., {Fan}, X., {et~al.} 2023, \href{http://dx.doi.org/10.3847/2041-8213/acc9c8}{\color{magenta}\apjl}, \href{https://ui.adsabs.harvard.edu/abs/2023ApJ...951L...5Y}{951, L5}

\bibitem[{{Yang} {et~al.}(2020){Yang}, {Wang}, {Fan}, {Hennawi}, {Davies}, {Yue}, {Banados}, {Wu}, {Venemans}, {Barth}, {Bian}, {Boutsia}, {Decarli}, {Farina}, {Green}, {Jiang}, {Li}, {Mazzucchelli}, \& {Walter}}]{yang2020}
{Yang}, J., {Wang}, F., {Fan}, X., {et~al.} 2020, \href{http://dx.doi.org/10.3847/2041-8213/ab9c26}{\color{magenta}\apjl}, \href{https://ui.adsabs.harvard.edu/abs/2020ApJ...897L..14Y}{897, L14}

\bibitem[{{Zhu} {et~al.}(2022){Zhu}, {Li}, {Li}, {Maji}, {Yajima}, {Schneider}, \& {Hernquist}}]{zhu2022}
{Zhu}, Q., {Li}, Y., {Li}, Y., {et~al.} 2022, \href{http://dx.doi.org/10.1093/mnras/stac1556}{\color{magenta}\mnras}, \href{https://ui.adsabs.harvard.edu/abs/2022MNRAS.514.5583Z}{514, 5583}

\end{thebibliography}
%\bsp
\label{lastpage}

\appendix
%\section{The extreme value statistics}
%\label{EVS}
%random variable drawn from an underlying distribution. The power of EVS is that it allows a test of the underlying cosmology from the observation of a single extreme object. Extreme value statistics provides the most extreme deviation from the median of a given probabilistic distribution \citep{gumbel1958statistics,kotz2000extreme} 

\section{Halo mass estimation for GN-z11}
In this Appendix, we compute the range of the allowed masses for the DM halos that host GN-z11, based on the SFR and $M_*$ estimates, and considering different $M_{\rm BH}-M_*$ relations. We find that the fiducial value  corresponds to $M_{\rm h}=4\times 10^{11}~\rm M_{\odot}$ .
\label{halognz11}
\subsection{Halo mass estimation from the star formation rate}
%To calculate the probable range of the halo mass for GN-z11, we consider the observational estimation of Star Formation Rate (SFR) $ = 21^{+22}_{-10} \ \mathrm{M}_\odot \  \mathrm{yr}^{-1}$ \citep{tacchella2023} which is calculated using NIRCam photometry assuming no light from the active galactic nuclei. 
%The simplest relation between SFR and DM halo mass can be written as Schmidt-type expression,
%\begin{equation}  
%\label{eq-SFR_Schmidt} 
%\mathrm{SFR} = \epsilon_{*} \ \frac{\Omega_{b}}{\Omega_{c}} \ \frac{M_{DM}}{t_{ff}}
%\end{equation}
%where $\Omega_{b}$ is baryon density, $\Omega_{c}$ is dark matter density,  $\epsilon_{*}$ is the star formation efficiency, $M_{DM}$ is the mass of the DM halo and $t_{ff} = \left( 4 \pi G \rho \right)^{-1/2}$ is the freefall time of gas in halos where $\rho$ is the  
To estimate the halo mass $M_h$ from the SFR, we adopted the relation proposed by \cite{ferrara2023},
\begin{equation}  
\label{eq-SFR_Mhalo_relation} 
\mathrm{SFR} = 22.7 \ \left( \frac{\epsilon_{\rm SF}}{0.01} \right) \ \left( \frac{1+z}{8} \right)^{3/2} \ \left( \frac{M_{\rm h}}{10^{12} \ \mathrm{M}_\odot} \right) \ \mathrm{M}_\odot \ \mathrm{yr}^{-1}
\end{equation}
where $\epsilon_{\rm SF}$ is star formation efficiency that depends on the supernova (SN) feedback as in \cite{dayal2014}:
%. To get the best possible agreement with the observed $M_{\rm UV}$ range from the SFR, we also included impact of supernova (SN) feedback as feedback can reduce $\epsilon_{*}$ resulting in suppression of star formation. We consider physically well motivated form proposed by \cite{dayal2014},
\begin{equation}  
\label{eq-eps_star} 
\epsilon_{\rm SF} = \epsilon_{0} \ \frac{v_{c}^{2}}{v_{c}^{2} + f_{w} \ v_{s}^{2}},
\end{equation}
where $f_{w}=0.1$ is the coupling efficiency of SN energy with gas, $v_{c}(M)$ is the halo circular velocity \citep[e.g.][]{barkana2001}, $\epsilon_{0}$ is fixed to 0.02 to be consistent with local galaxy measurements \citep{krumholz2017book}, $v_{s} = \sqrt{\nu \ E_{0}}$ is the characteristic velocity corresponding to the SN energy ($E_{0}$) released per unit stellar mass. Considering $E_{0} = 10^{51} \ \mathrm{erg}$ and $\nu^{-1} = 52.89 \ \mathrm{M}_\odot$ same as in \cite{ferrara2023}, we get $v_{s} = 975 \ \mathrm{km \ s^{-1}}$.  %We use \cite{barkana2001} for relation between circular velocity and total mass.

The SFR of GN-z11 derived from the NIRCam photometry (assuming no contribution from the AGN) is $ {\rm SFR} = 21^{+22}_{-10} \ \mathrm{M}_\odot \  \mathrm{yr}^{-1}$ \citep{tacchella2023}. This corresponds to $M_{\rm h}=4\times 10^{11}~\rm M_{\odot}$. By considering the 2 $\sigma$ deviation, we end up with the following possible range of halo mass: $5 \times 10^{10} < (M_{\rm h}/\mathrm{M}_\odot) < 1 \times 10^{12}$.

%$10.15 < \mathrm{log}_{10} (M_{\rm h}/\mathrm{M}_\odot) < 12.2$. 
%Using Eq. \ref{eq-SFR_Mhalo_relation} and Eq. \ref{eq-eps_star}, we calculate estimated lower and upper limit of halo mass corresponding to the given SFR. The halo mass corresponds to the SFR = 0.1 \mb{(Comment MB: Here we need to justify with physical reason about the lower limit as the constraint from Tacchella et al is $21_{-10}$ which can be negetive if we consider 3 sigma deviation. We can instead use 2 limit which gives lower bound of SFR=1 Msun)} is $ 10^{10.15} \mathrm{M}_\odot$, and for SFR = 87 is $ 10^{12.2} \mathrm{M}_\odot$. Thus, the estimated halo mass range would be $10.15 < \mathrm{log}_{10} (M_{\rm h}/\mathrm{M}_\odot) < 12.2$. 

\subsection{Halo mass estimation from the stellar mass}
Stellar mass $M_{*}$ can be related to halo mass $M_{\rm h}$ using the following relation:
\begin{equation}  
\label{eq-DM-Stellar_mass_relation} 
M_{\rm h} = \frac{\Omega_{\rm dm}}{\Omega_{\rm b}} \ \left( \frac{M_{*}}{\epsilon_{\rm *}} \right),
\end{equation}
where $\epsilon_{\rm *}=0.1$ is the conversion efficiency of baryons to stars. The stellar mass of GN-z11 derived from the NIRCam photometry is $\mathrm{log}_{10} (M_{*} / \mathrm{M}_\odot) = 9.1^{+0.3}_{-0.4}$ \citep{tacchella2023}. This corresponds to $M_{\rm h}=7\times 10^{10}~\rm M_{\odot}$. By considering the 3$\sigma$ deviation, we ended up with the following possible range of halo mass: $ 4 \times 10^9 < (M_{\rm h}/\mathrm{M}_\odot) < 5 \times 10^{11}$. 
%in the derived stellar mass value and fixed $\epsilon_{\rm st} = 0.1$ \mb{(Comment MB: Do we need a reference here?)}. The lower bound corresponds to $M_{*} = 10^{7.9} \mathrm{M}_\odot$ is $M_{\rm h} = 10^{9.7} \mathrm{M}_\odot$ and upper bound for the $M_{*} = 10^{10} \mathrm{M}_\odot$ is $M_{\rm h} = 10^{11.8} \mathrm{M}_\odot$. The estimated halo mass range from these calculation is $9.7 < \mathrm{log}_{10} (M_{\rm h}/\mathrm{M}_\odot) < 11.8$. 

%\subsubsection{Dark matter halo mass using scaling relation}
 %Considering the scaling relation between dark matter halo mass and black hole mass, we can estimate the dark matter halo mass. For $m=3,3.5,4$, the BH mass from the scaling relation is $M_{\rm BH} = $ 
\subsection{Halo mass estimation from the black hole mass}
In the local Universe, BH mass scales with stellar mass as $M_{\rm BH}\sim 10^{-4}~M_*$ \citep{reines2015}. However, it has been found that high-$z$ SMBHs can be overmassive with respect to their low-z counterparts by a factor greater than ten \citep{pensabene2020,pacucci2023}. In particular, \cite{pacucci2023} suggest the following relation for the high $z$: $M_{\rm BH}\sim 10^{-2}~M_*$. By combining the local and high-$z$ relations with Eq. \ref{eq-DM-Stellar_mass_relation} and assuming $M_{\rm BH} = 1.6 \times 10^6 \ \rm M_\odot$ \citep{maiolino2023a}, we estimate that GN-z11 should be hosted by a DM halo of mass $  9 \times 10^9 < (M_{\rm h}/\mathrm{M}_\odot) < 9 \times 10^{11}$.

%Using these estimated range we can exclude the "$b = 3$" value in Eq. \ref{eq-BH-DM_halo_mass} corresponds to $M_{\rm h} = 10^{9.2} \mathrm{M}_\odot$ which is smaller than the lower bounds calculated using SFR and stellar mass. To get the most conservative estimations, we have considered all the possible values of "$b$" except 3 in the scaling relation to calculate EVS PDF. 
%To get the most conservative estimations, we have taken least mass of the two calculations for the lower bound and highest mass of both for upper bound and calculated "N". Finally, we have integrated the Equation \ref{eq-EVS_N_GNz11} considering $m_{1} = 10^{9.62}$ and $m_{2} = 10^{12.2}$ resulting the N value to be 13988 ($\approx14000$) (NOTE TO BE ERASED : previous values 996.1 ($\approx1000$)). We have also checked the ''N" value considering higher DM halo mass in the upper limit of integration than the considered upper bound of $10^{12.2} M_\odot$. However it does not impact ''N" value as the slope of halo mass function above $10^{12.2} M_\odot$ is very steep. 
\end{document}